\documentclass[amsmath,amssymb,floatfix,aps,superscriptaddress,nofootinbib,notitlepage,twocolumn,pra]{revtex4-2}

\usepackage{amsmath,amsthm,amssymb,bm,graphicx,float,xcolor,changes,mathtools,enumitem}
\usepackage[colorlinks=true,urlcolor=purple,linkcolor=blue,citecolor=green,bookmarks=false]{hyperref}
\allowdisplaybreaks

\newcommand{\be}{\begin{equation}}
\newcommand{\ee}{\end{equation}}
\newcommand{\6}{\partial}
\newcommand{\inti}{\int_{-\infty}^{+\infty}}

\begin{document}

\title{Exact spectral function of the Tonks-Girardeau gas at finite temperature}

\author{Ovidiu I. P\^{a}\c{t}u}
\affiliation{Institute for Space Sciences, Bucharest-M\u{a}gurele, R 077125, Romania}

\begin{abstract}

We report on the derivation of determinant representations for the Green's functions and spectral
function of the trapped Tonks-Girardeau gas  on the lattice and in the continuum. Our results are valid
for any type of statistics of the constituent particles, at zero and finite temperature  and arbitrary
confining potentials, including nonequilibrium scenarios induced by sudden changes of the external potential.
In addition, they are also extremely efficient and easy to implement numerically with the main computational
effort being represented by  the calculation of partial overlaps of the dynamically evolved single particle wavefunctions.
In the lattice case we show that the spectral function of a system with a strong harmonic potential presents only
two singular lines compared with three singular lines in the case of a homogeneous system.

\end{abstract}

\maketitle

\onecolumngrid

\section{Introduction}\label{s0}

Due to the unprecedented degree of control over dimensionality, purity, strength of the interaction and statistics of the constituent
particles ultracold gases represent an versatile platform which allows for the investigation of many-body physics which would be very
difficult to study in solid state systems \cite{CCGO11,GBL13}. The experimental realization with ultracold atomic gases of many physical
systems which are well approximated by integrable or weakly broken integrable systems paved the way for the exploration of fundamental
theoretical questions regarding the long time dynamics and lack of thermalization in such systems \cite{MVBF22,MV22}.

A paradigmatic model which is now routinely realized in laboratories is the Lieb-Liniger (LL) model \cite{LL63} which describes one-dimensional (1D)
bosons with repulsive contact interactions. In the limit of infinite repulsion the system is in the so-called Tonks-Girardeau (TG) regime
\cite{Gira60,Olsh98,MSKE03,KWW04,PWMM04} which allows for a comprehensive analytical investigation of the correlation functions due to the
knowledge of the wavefunctions via the Bose-Fermi mapping \cite{Gira60}. The lattice counterpart of the Tonks-Girardeau gas is represented
by hard-core bosons on the lattice (the Bose-Hubbard model with infinite repulsion) or, equivalently, the isotropic XY spin-chain \cite{LSM61}
also known as the  XX spin-chain. The generalizations of the Lieb-Liniger and lattice Bose-Hubbard models to arbitrary statistics were introduced
in \cite{Kundu,BGO06,HZC09,KLMR11,BJEP21}  and   several proposals  of experimental realizations of such interesting  systems  with ultracold
atoms were proposed using Raman assisted tunneling \cite{KLMR11,GS15},  periodically driven lattices \cite{SSE16} or multicolor lattice-depth
modulation \cite{CGS16,GCS18}.

In the study of many-body systems the Green's functions and spectral function (the imaginary part of the retarded Green's function) are of
primary importance \cite{Mah00}. In particular, the spectral function provides fundamental information about the momentum distribution and
elementary excitations of the system. While in solid state systems the spectral function is usually accessed using angle-resolved photoemission
spectroscopy (ARPES) \cite{Dam04} in the case of ultracold gases it has been measured by radio-frequency spectroscopy \cite{CBAR04,VPDM18} and
its momentum-resolved extension \cite{SGJ08,DCG09}, Bragg spectroscopy \cite{VKDV08}, lattice modulation spectroscopy \cite{JSGME08,GTUJ11} and,
recently, an ARPES analogue has also been proposed \cite{BGDK18}.

In the case of systems of impenetrable particles in 1D knowledge of the wavefunctions, which can be obtained using the Bose-Fermi \cite{Gira60,YG05}
or Anyon-Fermi \cite{Girar,PKA1} mappings, opens the way for the derivation of determinant representations for the Green's functions. These
representations, as Toeplitz \cite{Gira60,Len64,BM71a,BM71b,MBA71,SSC07,SC08} or Hankel \cite{P03,FFGW03,FFGW04,MPC16,Hao16,HS17}  determinants for
homogenous or harmonically trapped finite size systems or as Fredholm determinants \cite{S63,Len66,KS90,PKA08,CIKT93,Pat15,GKSS19,ZNIG22} for
systems in the thermodynamic limit,  are extremely easy to implement numerically but they also represent the starting point for the derivation
of rigorous analytical results.
In general  the integral operators of the Fredholm determinants describing the correlators  of impenetrable particles are of a special type called
integrable  integral operators and due to their special structure they allow for the derivation of classical integrable differential equations for
the correlators  and the investigation of the asymptotics by solving an associated Riemann-Hilbert problem \cite{KBI,IIKS93,B91,GIK98,CZ04,P19}.
Other methods for deriving the asymptotics from the determinant representations are the use of Szeg\"o's theorem and Fisher-Hartwig conjecture
\cite{Len72,BM71a,BM71b,MBA71,FFGW03,SSC07,SC08,O07}, momentum space approach \cite{VT79}, connections with Painlev\'e transcendents
\cite{JMMS80,FFGW03,FFGW04}, the replica method \cite{Gan04,GS06,CS09,MPC16}, form factor expansions at zero  \cite{KKMS12,KT11,K15} and finite
temperature \cite{GKKK17,GKS20} or the effective form  factor approach \cite{GIZ21,ZNIG22,CG22}.

In this article we derive determinant representations for the space-, time-, and temperature-dependent Green's functions and spectral function for
a 1D system of impenetrable  particles of arbitrary statistics  in the presence of a confining potential.  These representations, for both  lattice
and continuum systems, are  also valid  in nonequilibrium scenarios due to rapid changes of the  confining potential or in the case in which the initial
state is not an eigenstate of the final Hamiltonian.
Our results, obtained via summation of the form factors, represent the generalization at finite temperature and arbitrary statistics of the representations
obtained  by Settino {\it et al.}  \cite{SGPM21} for  bosonic systems at zero temperature and  in nonequilibrium scenarios the equal time correlators reduce
to the results derived in \cite{PB07,AGBK17,P20}. Using an elegant operatorial method \cite{W22b} Wang derived in \cite{W22} similar equivalent representations but only in the case of
time-independent external potentials. In addition to being the starting point for the  analytical analysis of the asymptotic properties our determinant
representations have the advantage of being extremely easy and fast to implement numerically  with the main quantities which need to be computed being
the partial overlaps of the dynamically evolving single particle basis. This allows us to show numerically that while in the case of a homogeneous system
on the latice the  spectral function of a system of hard-core bosons presents three singular lines (the first two corresponding to the Type I
and Type II excitations in Lieb's classification \cite{L63} and the third one due to the presence of the lattice \cite{SGPM21}) the addition of a harmonic
potential has a significant effect on the relative weights and positions of each spectral line. As the curvature of the potential increases the spectral weight
of one of the lines diminishes approaching zero while a Mott insulator region develops in the center of the trap.

The plan of the paper is as follows. In Sect. \ref{s1} we introduce the anyonic TG gas in the continuum and present the eigenstates and their dynamics.
The determinant representation for the correlators at finite temperature obtained via summation of the form factors  is presented in Sect. \ref{s2}.
The lattice counterpart of the continuum TG gas is introduced in Sect. \ref{s3} and the  results for the lattice correlators can be found in Sect. \ref{s4}.
In Sect. \ref{s5} we compare our results  with previous representations derived in the literature in certain limits and in Sect. \ref{s6} we present
and discuss  numerical results for the spectral function of a lattice TG gas with harmonic trapping. We conclude in Sect. \ref{s7}. Some technical
details  regarding the equal time limit of one correlation function can be found in Appendix \ref{a1}.

\section{The anyonic Tonks-Girardeau gas in the continuum}\label{s1}

In this paper we will derive determinant representations for the correlators of 1D impenetrable particles with arbitrary statistics in the
continuum and on the lattice. We start with the case of continuum systems. A system of $N$  anyons in the continuum with repulsive contact
interactions in the presence of an external confining potential $V(x,t)$ is described by the Hamiltonian
\be\label{ham2}
\mathcal{H}=\int dx\,   \frac{\hbar^2}{2 m}(\6_x\Psi^\dagger)(\6_x\Psi)
+g \,\Psi^\dagger\Psi^\dagger\Psi\Psi+ (V(x,t)-\mu)\,\Psi^\dagger\Psi\, ,
\ee
where the anyonic fields $\Psi^\dagger(x), \Psi(x)$ satisfy the commutation relations
\begin{subequations}\label{comm}
\begin{align}
\Psi(x)\Psi^\dagger(y)&=-e^{- i \pi \kappa\, \mbox{\small{sign}}(x-y)}\Psi^\dagger(y)\Psi(x)+\delta(x-y)\, ,\\
\Psi(x)\Psi(y)&=-e^{i \pi \kappa\, \mbox{\small{sign}}(x-y)}\Psi(y)\Psi(x)\, ,
\end{align}
\end{subequations}
with $\kappa\in [0,1]$ the statistics parameter and $\mbox{{sign}}(x)=|x|/x \, ,\mbox{{sign}}(0)=0$. In (\ref{ham2}) $\hbar$ is the
reduced Planck constant, $m$ is the mass of the particles, $\mu$ the chemical potential and $g$  is the strength of the repulsive
interaction. The  commutation relations  (\ref{comm}) are fermionic at coinciding points, $x=y$, and, as we vary the statistics parameter,
they interpolate continuously between the bosonic commutation relations for $\kappa=1$ and the fermionic anticommutation relations for
$\kappa=0$. We will consider a particular type of time-dependent confining potentials $V(x,t)$ which describe certain quantum quenches.
More precisely we will consider scenarios in which at $t=0$ the system is described by $V(x,t=0)=V_I(x)$ and for $t>0$ we have
$V(x,t>0)=V_F(x)$. For example, the experimentally relevant situation of a system released from an harmonic trap is described by
$V(x,t=0)=m\omega_0^2 x^2/2$ and $V(x,t>0)=0$. We will denote the initial Hamiltonian by $\mathcal{H}_I$ and the final Hamiltonian
which governs the subsequent dynamics after the quench by $\mathcal{H}_F$. For time-independent Hamiltonians we obviously have
$\mathcal{H}_I=\mathcal{H}_F$.

In the absence of an external potential the Hamiltonian (\ref{ham2}) describes the integrable anyonic Lieb-Liniger model \cite{Kundu,
BGO06,PKA,BJEP21} which is the natural generalization to arbitrary statistics of the bosonic Lieb-Liniger model \cite{LL63}
 (see also \cite{KBI,FranB} and references therein). The realization that integrable and near-integrable
systems do not thermalize \cite{KWW06,RDO08,RDYO07} sparked renewed interest in the nonequilibrium dynamics of such systems. The dynamics
of the LL model in various nonequilibrium scenarios has been studied intensely in the last decade  see
\cite{GW00,DGW02,BBIS04,MG05,delC,BPG08,CSC13,KSCC13,KCC14,WRDK14,NWBC14,NC14,CGK15,BWEN16,BCS17,BB17,FUBS17,MFB18,CDDK19,SBDD19,RBD19,SS20,BLDN20,GE20,GE21,DCVA20,ASK20,BDD20,HP21,WMLZ20,MZLD21}.
 When $V(x,t)\ne 0$ the Hamiltonian (\ref{ham2}) is no longer integrable except when $g=0$ and in the impenetrable limit $g=\infty$ also known as the Tonks-Girardeau regime.
This regime of very strong repulsive interaction will be the main focus of this paper. At $t=0$ the eigenstates of the system described by $\mathcal{H}_I$ are \cite{KBI,PKA,P20}
\begin{align}\label{eigen}
|\boldsymbol{\psi}_N(\boldsymbol{k})\rangle=&\frac{1}{\sqrt{N!}}\int dx_1 \cdots dx_N\,  \psi_{N}(x_1,\cdots,x_N|\boldsymbol{k})
\Psi^\dagger(x_N)\cdots\Psi^\dagger(x_1)|0\rangle\, ,
\end{align}
where $|0\rangle$ is the Fock vacuum  satisfying  $\Psi(x)|0\rangle=\langle 0|\Psi^\dagger(x)=0$ for all $x$ and
$\langle 0|0\rangle=1$. Each eigenstate is indexed by a set of integers $\boldsymbol{k}=(k_1,\cdots,k_N)$ and the many-body anyonic wavefunction
is determined  using the  Anyon-Fermi mapping \cite{Girar,PKA1} which allows us to write it in terms of the wavefunction of noninteracting fermions
subjected to the same external potential
\be\label{wavef}
\psi_{N}(x_1,\cdots,x_N|\boldsymbol{k})=\frac{i^{N(N-1)/2}}{\sqrt{N!}}\prod_{1\le a<b\le N}e^{i\frac{\pi \kappa}{2}\,\mbox{\small{sign}}(x_a-x_b)}
\det\left[\phi_{k_a}(x_b)\right]_{a,b=1,\cdots,N}\, ,
\ee
with $\phi_k(x)$ the eigenfunctions of the initial  single-particle Hamiltonian $\mathcal{H}_{I}^{\mbox{\tiny{SP}}}(x)=-(\hbar^2/2m)(\6^2/\6 x^2)+V(x,t=0)$
i.e., $\mathcal{H}_{I}^{\mbox{\tiny{SP}}}(x)\phi_{k}(x)=\varepsilon(k)\phi_{k}(x)$ and $\varepsilon(k)$ the single particle dispersion relation.
Using $e^{i\frac{\pi \kappa}{2}\mbox{\small{sign}}(x_a-x_b)}=\cos (\pi\kappa/2)+i\, \mbox{sign}(x_a-x_b)\sin (\pi\kappa/2)$ it is easy to see
that  (\ref{wavef}) reproduces the bosonic result \cite{Gira60, Len66} when  $\kappa=1$ and  satisfies
\begin{align}\label{asymm}
\psi_{N}(\cdots,x_i,x_{i+1},\cdots|\boldsymbol{k})=&-e^{i\pi\kappa\, \mbox{\small{sign}}(x_i-x_{i+1})}
\psi_{N}(\cdots, x_{i+1},x_i,\cdots|\boldsymbol{k})\, .
\end{align}
being symmetric(anti-symmetric) under the  permutation of two particles when the system is bosonic $(\kappa=1)$ [fermionic $(\kappa=0)$].
For an anyonic system, $\kappa\in(0,1)$, Eq.~(\ref{asymm}) indicates that the   space-reversal symmetry is broken resulting   in a non-symmetric
momentum distribution. The eigenstates (\ref{eigen}) form a complete set, are normalized $\langle\boldsymbol{\psi}_N(\boldsymbol{k})
|\boldsymbol{\psi}_N(\boldsymbol{k'})\rangle=\delta_{\boldsymbol{k},\boldsymbol{k'}}$ and satisfy
\be
\mathcal{H}_I |\boldsymbol{\psi}_N(\boldsymbol{k})\rangle=E_N(\boldsymbol{k})|\boldsymbol{\psi}_N (\boldsymbol{k})\rangle\ \ \mbox{ with }
E_N(\boldsymbol{k})=\sum_{i=1}^N(\varepsilon(k_i)-\mu)\, .
\ee
In order to compute the form factors and, subsequently, the correlators we will also need
$|\boldsymbol{\psi}_N(\boldsymbol{k,t})\rangle=e^{- i t \mathcal{H}_F} |\boldsymbol{\psi}_N(\boldsymbol{k})\rangle$. In the case of a time-independent
external potential ($\mathcal{H}_I=\mathcal{H}_F$) the time-evolved eigenstate $|\boldsymbol{\psi}_N(\boldsymbol{k,t})\rangle$ is defined by a similar expression with (\ref{eigen})
with the time-dependent many-body wavefunction given by \cite{GW00,P20}
\be\label{waveft}
\psi_{N}(x_1,\cdots,x_N|\boldsymbol{k},t)=\frac{i^{N(N-1)/2}}{\sqrt{N!}}\prod_{1\le a<b\le N}e^{i\frac{\pi \kappa}{2}\,\mbox{\small{sign}}(x_a-x_b)}
\det\left[\phi_{k_a}(x_b,t)\right]_{a,b=1,\cdots,N}\, ,
\ee
where $\phi_k(x,t)=e^{-i \varepsilon(k) t}\phi_k(x)$. In the case of a quantum quench ($\mathcal{H}_I\ne\mathcal{H}_F$) the  many-body wavefunction is
also given by (\ref{waveft}) with $\phi_k(x,t)$ being the unique time-dependent solution of the Schr\"odinger equation
$i\hbar \6 \phi_{j}(x,t)/\6 t=H_F^{\mbox{\tiny{SP}}}(x) \phi_{j}(x,t)$ with $H_F^{\mbox{\tiny{SP}}}(x)=-(\hbar^2/2m)(\6^2/\6 x^2)+V(x,t>0)$ and
initial condition $\phi_k(x,0)=\phi_k(x)$ [remember that $\phi_k(x)$ are eigenfunctions of the initial single-particle Hamiltonian
$\mathcal{H}_{I}^{\mbox{\tiny{SP}}}(x)=-(\hbar^2/2m)(\6^2/\6 x^2)+V(x,t=0)$].

Finally, let us give some concrete examples of systems which can be considered:

\textit{Harmonic trapping:} $V_I(x)=m\omega^2 x^2/2$. In this case the single particle functions are the Hermite functions
\be\label{hermite}
\phi_k(x)=e^{- m\omega^2 x^2/2}\frac{1}{\sqrt{2^k k!}} \left(\frac{m \omega}{\pi}\right)^{1/4}H_k(\sqrt{m\omega}\, x)\, ,\ k=0,1,2,\cdots
\ee
with $\varepsilon(k)=\hbar\omega(k+1/2)$.

\textit{Triangular potential:} $V_I(x)=|x|$ (Chap. 8.1.2 of \cite{VS}). In this case for $k$ even $(2,4,\cdots)$ the single particle functions are
\be
\phi_k(x)=\left(\frac{2 m}{\hbar^2}\right)^{1/6}\frac{1}{\sqrt{-a_k'}Ai(a_k')}Ai\left[\left(\frac{2 m}
{\hbar^2}\right)^{1/3}(|x|-\varepsilon(k))\right]\, , \  \varepsilon(k)=-a_{k+1}'\left(\frac{\hbar^2}{2m}\right)^{1/3}\, ,
\ee
while for $k$ odd $(1,3,\cdots)$ the functions are
\be
\phi_k(x)=\mbox{sign}(x)\left(\frac{2 m}{\hbar^2}\right)^{1/6}\frac{1}{Ai'(a_k)}Ai\left[\left(\frac{2 m}
{\hbar^2}\right)^{1/3}(|x|-\varepsilon(k))\right]\, , \  \varepsilon(k)=-a_{k+1}\left(\frac{\hbar^2}{2m}\right)^{1/3}\, ,
\ee
where $Ai(x)=\frac{1}{2\pi}\inti e^{i(z^3/3+x z)}\, dz$ is the Airy function and $a_k$ and $a_k'$ are the k-th zeros of $Ai(x)$ and
 $Ai'(x)$ respectively.

\textit{Dirichlet boundary conditions:} The single particle functions satisfy $\phi_k(0)=\phi_k(L)=0$ (the system is defined on $[0,L]$)
and are given by
\be
\phi_k(x)=\sqrt{\frac{2}{L}}\sin\left(\frac{\pi k x}{L}\right)\, , k=1,2,\cdots\, , \ \ \ \
\varepsilon(k)=\frac{\hbar^2}{2m}\frac{\pi^2k^2}{L^2}\, .
\ee

\textit{Neumann boundary conditions:} The single particle functions satisfy $\frac{d \phi_k}{d x}(0)=\frac{d \phi_k}{d x}(L)=0$ (the
system is defined on $[0,L]$) and are given by
\be
\phi_k(x)=\left\{\begin{array} {ll} \frac{1}{\sqrt{L}}    & k=0\\
                                           \frac{2}{\sqrt{L}}\cos\left(\frac{\pi k x}{L}\right)\, & k=1,2,\cdots\, ,
                                           \end{array}\right.
                                           \varepsilon(k)=\frac{\hbar^2}{2m}\frac{\pi^2k^2}{L^2}\, .
\ee
We should point out that our formalism is not valid for systems with no confining potential and periodic boundary conditions. In this case
determinant representations for the correlators  can be found in \cite{Len64,Len66,KS90,KBI,SSC07,SC08,PKA08}.

\section{Determinant representations for finite temperature correlators in the continuum }\label{s2}

In this section we will derive determinant representations for the space-, time-, and temperature-dependent field-field correlators of impenetrable
particles in the continuum which can be investigated numerically or constitute the starting point for the analytic investigation of their asymptotic behaviour.
More precisely, we are interested in
\begin{align}\label{gm}
g^{(-)}(x,t;y,t')&=\langle\Psi^\dagger(x,t)\Psi(y,t')\rangle_{\mu,T}\, , \nonumber\\
&=\mbox{Tr} \left[e^{-\mathcal{H}_I/T}\Psi^\dagger(x,t)\Psi(y,t')\right]/\mbox{Tr} \left[e^{-\mathcal{H}_I/T}\right]\, , \nonumber\\
&=\sum_{N=0}^\infty\, \sum_{k_1<\cdots<k_{N+1}}e^{-E_{N+1}(\boldsymbol{k})/T} \langle\boldsymbol{\psi}_{N+1}(\boldsymbol{k})|\Psi^\dagger(x,t)\Psi(y,t')|\boldsymbol{\psi}_{N+1}(\boldsymbol{k})\rangle/
\sum_{N=0}^\infty\, \sum_{k_1<\cdots<k_{N}}e^{-E_{N}(\boldsymbol{k})/T}\, ,
\end{align}
and
\begin{align}\label{gp}
g^{(+)}(x,t;y,t')&=\langle\Psi(x,t)\Psi^\dagger(y,t')\rangle_{\mu,T}\, , \nonumber\\
&=\mbox{Tr} \left[e^{-\mathcal{H}_I/T}\Psi(x,t)\Psi^\dagger(y,t')\right]/\mbox{Tr} \left[e^{-\mathcal{H}_I/T}\right]\, , \nonumber\\
&=\sum_{N=0}^\infty\, \sum_{q_1<\cdots<q_{N}}e^{-E_{N}(\boldsymbol{k})/T} \langle\boldsymbol{\psi}_{N}(\boldsymbol{q})|\Psi(x,t)\Psi^\dagger(y,t')|\boldsymbol{\psi}_{N}(\boldsymbol{q})\rangle/
\sum_{N=0}^\infty\, \sum_{q_1<\cdots<q_{N}}e^{-E_{N}(\boldsymbol{q})/T}\, ,
\end{align}
where $\Psi^\dagger(x,t)=e^{ i \mathcal{H}_F t}\Psi^\dagger(x)e^{- i \mathcal{H}_F t}$ and $\Psi(x,t)=e^{ i \mathcal{H}_F t}\Psi(x)e^{- i \mathcal{H}_F t}$. These are the field-field correlators
evaluated in a thermal state of the initial Hamiltonian $\mathcal{H}_I$ described by the grandcanonical ensemble at temperature $T$ and chemical potential $\mu$. In equilibrium we have
$\mathcal{H}_I=\mathcal{H}_F$ while in a quench scenario  the subsequent time evolution of the fields is described by the final Hamiltonian $\mathcal{H}_F$. The field correlators (\ref{gm})
and (\ref{gp}) allow us to describe the usual six Green's functions (advanced, retarded, time-ordered, anti-time-ordered, greater, lesser) usually employed in the study of many-body systems
(see Chap. 2.9.1 of \cite{Mah00}). In particular, the greater and lesser Green's functions  are given by
\begin{align}
G^{>}(x,t;y,t')&=-i\langle \Psi(x,t)\Psi^\dagger(y,t')\rangle_{\mu,T}=-ig^{(+)}(x,t;y,t')\, ,\\
G^{<}(x,t;y,t')&=-i\langle \Psi^\dagger(y,t')\Psi(x,t)\rangle_{\mu,T}=-ig^{(-)}(y,t';x,t)\, ,
\end{align}
and the retarded Green's function is
\begin{align}\label{spectralfc}
G^{R}(x,t;y,t')=&\,\Theta(t-t')\left[G^>(x,t;y,t')+G^<(x,t;y,t')\right]\, ,\\
=&-i\,\Theta(t-t')\left[g^{(+)}(x,t;y,t')+g^{(-)}(y,t';x,t)\right]\, ,
\end{align}
where $\Theta(t)$ is the Heaviside function. The spectral function is the imaginary part of the Fourier transform of the retarded Green's function
\begin{align}
A(k,\omega)=-\frac{1}{\pi} \mbox{Im}\, G^R(k,\omega)\, ,
\end{align}
with
\begin{align}
G^R(k,\omega)=\inti \, dt\, e^{i \omega t} \inti\inti\, dx dy\, e^{-i k(x-y)} G^R(x,t;y,0)\, .
\end{align}
The density  and the momentum distribution of the system are given by  $\rho(x,t)=g^{(-)}(x,t;x,t)$ and $n(k,t)=\inti\inti\, dx dy\, e^{-i k(x-y)} g^{(-)}(x,t;y,t)$, respectively.

\subsection{Form factors}

In order to derive determinant representations for the field correlators (\ref{gm}) and (\ref{gp}) we are going to employ the summation of form factors \cite{KS90,KBI}.
We are going to compute first the form factors for a finite size system and express the mean value of bilocal operators in an arbitrary state as a sum over them.
The summation can be performed using a well known technique known as the ``insertion of summation under the determinant''\cite{KS90,KBI} (this can be understood as
an application of a  slightly modified Cauchy-Binet formula \cite{SGPM21}) which will allow the derivation of a
single determinant for the mean value. The final result valid in  the
thermodynamic limit is obtained using von Koch's formula.

We start with the computation of the form factors for a finite size system. Inserting a resolution of the identity in each mean value of bilocal operators appearing in
(\ref{gm}) and (\ref{gp}) we obtain (the bar denotes complex conjugation)
\begin{align}
\langle\boldsymbol{\psi}_{N+1}(\boldsymbol{k})|\Psi^\dagger(x,t)\Psi(y,t')|\boldsymbol{\psi}_{N+1}(\boldsymbol{k})\rangle&=
\sum_{q_1<\cdots<q_N} \underbrace{\langle\boldsymbol{\psi}_{N+1}(\boldsymbol{k})|\Psi^\dagger(x,t)|\boldsymbol{\psi}_{N}(\boldsymbol{q})\rangle}_{\overline{F}_N(\boldsymbol{k},\boldsymbol{q}|x,t)}
\underbrace{\langle\boldsymbol{\psi}_{N}(\boldsymbol{q})|\Psi(y,t')|\boldsymbol{\psi}_{N+1}(\boldsymbol{k})\rangle}_{F_N(\boldsymbol{k},\boldsymbol{q}|y,t')}\, ,\label{bminus}\\
\langle\boldsymbol{\psi}_{N}(\boldsymbol{q})|\Psi(x,t)\Psi^\dagger(y,t')|\boldsymbol{\psi}_{N}(\boldsymbol{q})\rangle&=
\sum_{k_1<\cdots<k_{N+1}} \underbrace{\langle\boldsymbol{\psi}_{N}(\boldsymbol{q})|\Psi(x,t)|\boldsymbol{\psi}_{N+1}(\boldsymbol{k})\rangle}_{F_N(\boldsymbol{k},\boldsymbol{q}|x,t)}
\underbrace{\langle\boldsymbol{\psi}_{N+1}(\boldsymbol{k})|\Psi^\dagger(y,t')|\boldsymbol{\psi}_{N}(\boldsymbol{q})\rangle}_{\overline{F}_N(\boldsymbol{k},\boldsymbol{q}|y,t')}\, ,\label{bplus}
\end{align}
which shows that each of this quantities can be expressed as sums over form factors which are defined as
\begin{align}
F_{N}(\boldsymbol{k},\boldsymbol{q}|x,t)=\langle\boldsymbol{\psi}_{N}(\boldsymbol{q})|\Psi(x,t)|\boldsymbol{\psi}_{N+1}(\boldsymbol{k})\rangle\, ,
\end{align}
where $\boldsymbol{k}=(k_1,\cdots,k_{N+1})$ and $\boldsymbol{q}=(q_1,\cdots,q_N)$ describe arbitrary states with $N+1$ and $N$ particles, respectively.
Note that we need to define only the form factor of the $\Psi(x,t)$ operator, the equivalent quantity for the creation operator  $\Psi^\dagger(x,t)$ is given
by the complex conjugate of $F_{N}(\boldsymbol{k},\boldsymbol{q}|x,t)$. Using the equal-time commutation relations (\ref{comm}) and the definition of the
eigenstates (\ref{eigen}) we obtain
\be\label{i1}
F_{N}(\boldsymbol{k},\boldsymbol{q}|x,t)=e^{i\mu t} \sqrt{N+1}\int_{L_-}^{L_+} dx_1\cdots dx_N \psi_{N+1}(x_1,\cdots,x_N,x|\boldsymbol{k},t)
\overline{\psi}_{N}(x_1,\cdots,x_N|\boldsymbol{q},t)\, ,
\ee
with
\begin{align}
\psi_{N+1}(x_1,\cdots,x_N,x|\boldsymbol{k},t)&=\frac{i^{N(N+1)/2}}{\sqrt{(N+1)!}}\left(\prod_{1\le a<b\le N}e^{i\frac{\pi\kappa}{2}\mbox{\small{sign}}(x_a-x_b)}\right)
\left(\prod_{j=1}^N e^{i\frac{\pi\kappa}{2}\mbox{\small{sign}}(x_j-x)}\right)\nonumber\\
&\qquad\qquad\qquad \qquad\times \sum_{P\in S_{N+1}}(-1)^P\prod_{j=1}^N\phi_{k_P(j)}(x_j,t)\phi_{k_{P(N+1)}}(x,t)\, ,\label{wave1}\\
\overline{\psi}_{N}(x_1,\cdots,x_N,|\boldsymbol{q},t)&=\frac{(-i)^{N(N-1)/2}}{\sqrt{N!}}\left(\prod_{1\le a<b\le N}e^{-i\frac{\pi\kappa}{2}\mbox{\small{sign}}(x_a-x_b)}\right)
\sum_{Q\in S_{N}}(-1)^Q\prod_{j=1}^N\overline{\phi}_{q_Q(j)}(x_j,t)\, ,\label{wave2}
\end{align}
where $S_{N+1}$ is the group of permutations of $N+1$ elements, $(-1)^P$ the signature of the permutation and $\phi_k(x,t)$ are the time-evolved
single particle eigenfunctions (when we do not have a quench we have  $\phi_k(x,t)=e^{-i\varepsilon(k)t}\phi_k(x)$).
In (\ref{i1}) $L_{\pm}$  quantify the size of the system which depends on the potential or boundary conditions. For example in the case of a harmonic potential
we have $L_\pm=\pm\infty$ while in the case of a system with Dirichlet boundary conditions at $0$  and $L$ we have $L_-=0$ and $L_+=L$. In all cases due to the
completeness and orthonormality of the single particle eigenfunctions we have
\be\label{complete}
\int_{L_-}^{L_+}\phi_k(v,t)\overline{\phi}_q(v,t)\, dv=\delta_{k,q}\, .
\ee
Using the explicit form of the wavefunctions (\ref{wave1}) and (\ref{wave2}) the analytic expression for the form factor (\ref{i1}) can be
written as a sum of factorized terms
\begin{align}
F_{N}(\boldsymbol{k},\boldsymbol{q}|x,t)=\frac{e^{i\mu t} i^N}{N!}\int_{L_-}^{L_+} dx_1\cdots dx_N\sum_{P\in S_{N+1}}\sum_{Q\in S_N} (-1)^{P+Q} \left(\prod_{j=1}^N e^{i\frac{\pi\kappa}{2}\mbox{\small{sign}}(x_j-x)}
\phi_{k_{P(j)}}(x_j,t)\overline{\phi}_{q_{Q(j)}}(x_j,t)\right)\phi_{k_{P(N+1)}}(x,t)\, ,
\end{align}
with
\begin{align}
\int_{L_-}^{L_+}e^{i\frac{\pi \kappa}{2}\mbox{\small{sign}}(v-x)}\phi_k(v,t)\overline{\phi}_q(v, t)\, dv&=e^{-i\frac{\pi\kappa}{2}}\int_{L_-}^x \phi_k(v,t)\overline{\phi}_q(v, t)\, dv+
e^{i\frac{\pi\kappa}{2}}\int_{x}^{L_+} \phi_k(v,t)\overline{\phi}_q(v, t)\, dv\, ,\nonumber\\
&=e^{-i\frac{\pi\kappa}{2}}\left(\delta_{k,q}-(1-e^{i\pi\kappa})\int_{x}^{L_+} \phi_k(v,t)\overline{\phi}_q(v, t)\, dv \right)\, ,
\end{align}
where we have used the completeness relation (\ref{complete}). Introducing
\be\label{deff}
f(k,q|\kappa,x,t)=\delta_{k,q}-(1-e^{i\pi\kappa})\int_{x}^{L_+} \phi_k(v,t)\overline{\phi}_q(v, t)\, dv \, ,
\ee
we find
\begin{align}
F_{N}(\boldsymbol{k},\boldsymbol{q}|x,t)&=\frac{e^{i\mu t}\left(i e^{-i\frac{\pi\kappa}{2}}\right)^N}{N!}\sum_{P\in S_{N+1}}\sum_{Q\in S_N} (-1)^{P+Q}
\left(\prod_{j=1}^N f(k_{P(j)},q_{Q(j)}|\kappa,x,t)\right)\phi_{k_{P(N+1)}}(x,t)\, ,\nonumber\\
&=\frac{e^{i\mu t}\left(i e^{-i\frac{\pi\kappa}{2}}\right)^N}{N!}\sum_{Q\in S_N} (-1)^{Q}
\left|
\begin{array}{cccc}
f(k_1,q_{Q(1)}|\kappa) &\cdots & f(k_1,q_{Q(N)}|\kappa) & \phi_{k_1}\\
\vdots  &\ddots  &\vdots  & \vdots\\
f(k_{N+1},q_{Q(1)}|\kappa) &\cdots & f(k_{N+1},q_{Q(N)}|\kappa) & \phi_{k_{N+1}}
\end{array}
\right|(x,t)\, .
\end{align}
For each permutation $Q$ rearranging the columns of the matrix appearing in the last expression $Q\rightarrow (1,\cdots,N)$
gives a $(-1)^Q$ factor and for each of the $N!$ permutation we obtain the same result. Therefore, the final result for the
form factor is
\be\label{ff}
F_{N}(\boldsymbol{k},\boldsymbol{q}|x,t)=e^{i\mu t}\left(i e^{-i\frac{\pi\kappa}{2}}\right)^N\, \det_{N+1} B(\boldsymbol{k},\boldsymbol{q}|x,t)\, ,
\ee
where $B$ is a square matrix of dimension $N+1$ with elements
\be\label{defb}
B_{ab}(\boldsymbol{k},\boldsymbol{q}|x,t)=\left\{\begin{array}{ll} f(k_a,q_b|\kappa,x,t)\, ,\ \ &a=1,\cdots,N+1\, ,\ \ b=1,\cdots,N\, ,\\
                               \phi_{k_a}(x,t)\, ,\ \  &a=1,\cdots,N+1\, ,\ \ b=N+1\, .\\
                               \end{array}
                               \right.
\ee
%

\subsection{Determinant representation for $\langle\boldsymbol{\psi}_{N+1}(\boldsymbol{k})|\Psi^\dagger(x,t)\Psi(y,t')|\boldsymbol{\psi}_{N+1}(\boldsymbol{k})\rangle$}

We can obtain a determinant representation for the mean values of bilocal operators appearing in the definition
of the correlation functions (\ref{gm}) and (\ref{gp}) by using the explicit formula for the form factors (\ref{ff})
and the ``insertion of summation under the determinant'' \cite{KS90,KBI}. From (\ref{bminus}) we have
\begin{align}
A^{(-)}&\equiv \langle\boldsymbol{\psi}_{N+1}(\boldsymbol{k})|\Psi^\dagger(x,t)\Psi(y,t')|\boldsymbol{\psi}_{N+1}(\boldsymbol{k})\rangle
         =\sum_{q_1<\cdots<q_N}\overline{F}_N(\boldsymbol{k},\boldsymbol{q}|x,t)F_{N}(\boldsymbol{k},\boldsymbol{q}|y,t')\, ,\nonumber\\
         &=e^{-i\mu(t-t')}\sum_{q_1<\cdots<q_N} \overline{\det_{N+1} B(\boldsymbol{k},\boldsymbol{q}|x,t)}\det_{N+1} B(\boldsymbol{k},\boldsymbol{q}|y,t')\, .
\end{align}
The product of the two determinants is symmetric in $q$'s and vanishes when two of them are equal, therefore we can write
\begin{align}\label{i3}
A^{(-)}&=\frac{e^{-i\mu(t-t')}}{N!}\sum_{q_1=1}^\infty\cdots \sum_{q_N=1}^\infty\overline{\det_{N+1} B(\boldsymbol{k},\boldsymbol{q}|x,t)}\det_{N+1} B(\boldsymbol{k},\boldsymbol{q}|y,t')\, ,\nonumber\\
&=\frac{e^{-i\mu(t-t')}}{N!}\sum_{q_1=1}^\infty\cdots \sum_{q_N=1}^\infty\sum_{P\in S_{N+1}}\sum_{Q\in S_{N+1}}(-1)^{P+Q}
\left(\prod_{j=1}^N \overline{f}(k_{P(j)},q_j|\kappa,x,t) f(k_{Q(j)},q_j|\kappa,y,t')\right)\nonumber\\
&\qquad\qquad\qquad\qquad\qquad\qquad\qquad\qquad\qquad\qquad\qquad\qquad\qquad\qquad\qquad\qquad\times\overline{\phi}_{k_{P(N+1)}}(x,t)\phi_{k_{Q(N+1)}}(y,t')\, ,\nonumber\\
&=\frac{e^{-i\mu(t-t')}}{N!}\sum_{q_1=1}^\infty\cdots \sum_{q_N=1}^\infty\sum_{R\in S_{N+1}}\sum_{Q\in S_{N+1}}(-1)^{R}
\left(\prod_{j=1}^N \overline{f}(k_{RQ(j)},q_j|\kappa,x,t) f(k_{Q(j)},q_j|\kappa,y,t')\right)\nonumber\\
&\qquad\qquad\qquad\qquad\qquad\qquad\qquad\qquad\qquad\qquad\qquad\qquad\qquad\qquad\qquad\qquad\times\overline{\phi}_{k_{RQ(N+1)}}(x,t)\phi_{k_{Q(N+1)}}(y,t')\, .
\end{align}
In the third line of (\ref{i3}) we have used the fact that for any two permutations $P$  and $Q$ we have $P=RQ$ with $R$ another permutation. It is easy to see that the
last line of (\ref{i3}) can be written as
\begin{align*}
A^{(-)}&=\frac{e^{-i\mu(t-t')}}{N!}\sum_{q_1=1}^\infty\cdots \sum_{q_N=1}^\infty\sum_{Q\in S_{N+1}}\nonumber\\
&\times\left|
\begin{array}{cccc}
\overline{f}(k_{Q(1)},q_1|x,t) f(k_{Q(1)},q_1|y,t') &\cdots &\overline{f}(k_{Q(1)},q_N|x,t) f(k_{Q(1)},q_N|y,t') &\overline{\phi}_{k_{Q(1)}}(x,t)\phi_{k_{Q(N+1)}}(y,t')\\
\vdots&\ddots&\vdots&\vdots\\
\overline{f}(k_{Q(N+1)},q_1|x,t) f(k_{Q(1)},q_1|y,t') &\cdots &\overline{f}(k_{Q(N+1)},q_N|x,t) f(k_{Q(1)},q_N|y,t') &\overline{\phi}_{k_{Q(N+1)}}(x,t)\phi_{k_{Q(N+1)}}(y,t')\\
\end{array}
\right|\, .
\end{align*}
Because $q_i$ appears only in the $i$-th column we can sum inside the determinant. Introducing two matrices (depending on the state $\boldsymbol{k}$)
\begin{align}
\widetilde{U}_{a,b}^{(-)}(x,t;y,t')&=\sum_{q=1}^\infty\overline{f}(k_a,q|\kappa,x,t)f(k_b,q|\kappa,y,t')\, ,\qquad a,b=1,\cdots, N+1\, ,\label{defum}\\
\widetilde{R}_{a,b}^{(-)}(x,t;y,t')&=\overline{\phi}_{k_a}(x,t)\phi_{k_b}(y,t')\, ,\qquad a,b=1,\cdots, N+1\, ,\label{defrm}
\end{align}
the last relation can be written as
\begin{align}
A^{(-)}&=\frac{e^{-i\mu(t-t')}}{N!}\sum_{Q\in S_{N+1}}
\left|
\begin{array}{cccc}
\widetilde{U}_{Q(1),Q(1)}^{(-)} &\cdots &\widetilde{U}_{Q(1),Q(N)}^{(-)} &\widetilde{R}_{Q(1),Q(N+1)}^{(-)}\\
\vdots&\ddots&\vdots&\vdots\\
\widetilde{U}_{Q(N+1),Q(1)}^{(-)} &\cdots &\widetilde{U}_{Q(N+1),Q(N)}^{(-)} &\widetilde{R}_{Q(N+1),Q(N+1)}^{(-)}\\
\end{array}
\right|(x,t;y,t')\, ,\nonumber\\
&=e^{-i\mu(t-t')}\sum_{j=1}^{N+1}\left|
\begin{array}{ccccc}
\widetilde{U}_{1,1}^{(-)} &\cdots &\widetilde{R}_{1,j}^{(-)} &\cdots & \widetilde{U}_{1,N+1}^{(-)}\\
\vdots&\ddots&\vdots&\ddots&\vdots\\
\widetilde{U}_{N+1,1}^{(-)} &\cdots &\widetilde{R}_{N+1,j}^{(-)} &\cdots & \widetilde{U}_{N+1,N+1}^{(-)}\\
\end{array}
\right|(x,t;y,t')
\end{align}
where in the last line we have reorganized the columns and rows such that $Q\rightarrow (1,\cdots,N+1)$. Using the fact that
 $\widetilde{R}_{a,b}^{(-)}$ is a rank 1 matrix we can write the final result
\begin{align}\label{bilocalm}
\langle\boldsymbol{\psi}_{N+1}(\boldsymbol{k})|\Psi^\dagger(x,t)\Psi(y,t')|\boldsymbol{\psi}_{N+1}(\boldsymbol{k})\rangle&=
e^{-i\mu(t-t')}\frac{\6}{\6z}\det_{N+1}\left.\left(\widetilde{U}^{(-)}+z\widetilde{R}^{(-)}\right)\right|_{z=0}\, ,\\
&=e^{-i\mu(t-t')}\left[\det_{N+1}\left(\widetilde{U}^{(-)}+\widetilde{R}^{(-)}\right)-\det_{N+1}\widetilde{U}^{(-)}\right]\, .
\end{align}

\subsection{Determinant representation for $\langle\boldsymbol{\psi}_{N}(\boldsymbol{q})|\Psi(x,t)\Psi^\dagger(y,t')|\boldsymbol{\psi}_{N}(\boldsymbol{q})\rangle$}

Similar to the previous case we can obtain a determinant formula for the mean value of  bilocal operators appearing in (\ref{gp}).  From (\ref{bplus}) we
have
\begin{align}
A^{(+)}&\equiv\langle\boldsymbol{\psi}_{N}(\boldsymbol{q})|\Psi(x,t)\Psi^\dagger(y,t')|\boldsymbol{\psi}_{N}(\boldsymbol{q})\rangle=
\sum_{k_1<\cdots<k_{N+1}} F_{N}(\boldsymbol{k},\boldsymbol{q}|x,t)\overline{F}_{N}(\boldsymbol{k},\boldsymbol{q}|y,t')\, ,\nonumber\\
&=e^{i\mu(t-t')}\sum_{k_1<\cdots<k_{N+1}} \det_{N+1}B(\boldsymbol{k},\boldsymbol{q}|x,t) \overline{\det_{N+1}B(\boldsymbol{k},\boldsymbol{q}|y,t')}\, .
\end{align}
The product of determinants is symmetric in $k$'s and vanishes when two of them are equal. Therefore the summation over all the states
with $N+1$ particles can be written as
\begin{align}\label{i4}
A^{(+)}&=\frac{e^{i\mu(t-t')}}{(N+1)!}\sum_{k_1=1}^\infty\cdots\sum_{k_{N+1}=1}^\infty\sum_{P\in S_{N+1}}\sum_{Q\in S_{N+1}}(-1)^{P+Q}
\left(\prod_{j=1}^N f(k_{P(j)},q_j|\kappa,x,t)\overline{f}(k_{Q(j)},q_j|\kappa,y,t')\right)\nonumber\\
&\qquad\qquad\qquad\qquad\qquad\qquad\qquad\qquad\qquad\qquad\qquad\qquad\qquad\qquad\qquad\qquad\times
\phi_{k_{P(N+1)}}(x,t)\overline{\phi}_{k_{Q(N+1)}}(y,t')\, ,\nonumber\\
&=\frac{e^{i\mu(t-t')}}{(N+1)!}\sum_{k_1=1}^\infty\cdots\sum_{k_{N+1}=1}^\infty\sum_{R\in S_{N+1}}\sum_{Q\in S_{N+1}}(-1)^{R}
\left(\prod_{j=1}^N f(k_{RQ(j)},q_j|\kappa,x,t)\overline{f}(k_{Q(j)},q_j|\kappa,y,t')\right)\nonumber\\
&\qquad\qquad\qquad\qquad\qquad\qquad\qquad\qquad\qquad\qquad\qquad\qquad\qquad\qquad\qquad\qquad\times
\phi_{k_{RQ(N+1)}}(x,t)\overline{\phi}_{k_{Q(N+1)}}(y,t')\, ,\nonumber\\
&=\frac{e^{i\mu(t-t')}}{(N+1)!}\sum_{k_1=1}^\infty\cdots\sum_{k_{N+1}=1}^\infty\sum_{Q\in S_{N+1}}\left(\prod_{j=1}^N\overline{f}(k_{Q(j)},q_j|\kappa,y,t')\overline{\phi}_{k_{Q(N+1)}}(y,t')\right)\nonumber\\
&\qquad\qquad\qquad\qquad\qquad\qquad\qquad\times\left|
\begin{array}{cccc}
f(k_{Q(1)},q_1|\kappa,x,t)&\cdots&f(k_{Q(1)},q_N|\kappa,x,t) &\phi_{k_{Q(1)}}(x,t)\\
\vdots&\ddots&\vdots&\vdots\\
f(k_{Q(N+1)},q_1|\kappa,x,t)&\cdots&f(k_{Q(N+1)},q_N|\kappa,x,t) &\phi_{k_{Q(N+1)}}(x,t)
\end{array}
\right|\, .
\end{align}
Like in the previous case  in the second line we have expressed
the permutations $P$ as a product of $R$ and $Q$. Now we multiply the $j$-th row of the determinant from the third line of (\ref{i4}) with
$\overline{f}(k_{Q(j)},q_j|\kappa,y,t')$ and the $N+1$-th row with $\overline{\phi}_{k_{Q(N+1)}}(y,t')$ obtaining
\begin{align*}
A^{(+)}&=
\frac{e^{i\mu(t-t')}}{(N+1)!}\sum_{k_1=1}^\infty\cdots\sum_{k_{N+1}=1}^\infty\sum_{Q\in S_{N+1}}\nonumber\\
&\times\left|
\begin{array}{cccc}
f(k_{Q(1)},q_1|x,t)\overline{f}(k_{Q(1)},q_1|y,t') & \cdots & f(k_{Q(1)},q_N|x,t)\overline{f}(k_{Q(1)},q_1|y,t') &\phi_{k_{Q(1)}}(x,t)\overline{f}(k_{Q(1)},q_1|y,t')\\
\vdots&\ddots&\vdots&\vdots\\
f(k_{Q(N)},q_1|x,t)\overline{f}(k_{Q(N)},q_N|y,t') & \cdots & f(k_{Q(N)},q_N|x,t)\overline{f}(k_{Q(N)},q_N|y,t') &\phi_{k_{Q(N)}}(x,t)\overline{f}(k_{Q(N)},q_N|y,t')\\
f(k_{Q(N+1)},q_1|x,t)\overline{\phi}_{k_{Q(N+1)}}(y,t') & \cdots & f(k_{Q(N+1)},q_N|x,t)\overline{\phi}_{k_{Q(N+1)}}(y,t') &\phi_{k_{Q(N+1)}}(x,t)\overline{\phi}_{k_{Q(N+1)}}(y,t')\\
\end{array}
\right|\, .
\end{align*}
In the previous expression $k_{Q(j)}$ appears only in the $j$-th row and, therefore, we can sum inside the determinant. Introducing
the $\boldsymbol{q}=(q_1,\cdots,q_n)$ dependent matrix and functions
\begin{align}
\widetilde{U}_{a,b}^{(+)}(x,t;y,t')&=\sum_{k=1}^\infty f(k,q_b|\kappa,x,t)\overline{f}(k,q_a|\kappa,y,t')\, , \qquad a,b=1,\cdots, N\, ,\label{defup}\\
\widetilde{e}_a(x,t;y,t')&=\sum_{k=1}^\infty f(k,q_a|\kappa,x,t)\overline{\phi}_k(y,t')\, ,\qquad a=1,\cdots, N\, ,\label{defe}\\
\widetilde{\overline{e}}_a(x,t;y,t')&=\sum_{k=1}^\infty \overline{f}(k,q_a|\kappa,y,t')\phi_k(x,t)\, ,\qquad a=1,\cdots, N\, ,\label{defebar}
\end{align}
together with
\be\label{defg}
g(x,t;y,t')=\sum_{k=1}^\infty\phi_k(x,t)\overline{\phi}_k(y,t')\, ,
\ee
the last expression can be written as (the sum over the $Q$ permutations produces $(N+1)!$ identical terms)
\begin{align*}
A^{(+)}&=
e^{i\mu(t-t')}
\left|
\begin{array}{cccc}
\widetilde{U}_{1,1}^{(+)} & \cdots & \widetilde{U}_{1,N}^{(+)} &\widetilde{\overline{e}}_1\\
\vdots&\ddots&\vdots&\vdots\\
\widetilde{U}_{N,1}^{(+)} & \cdots & \widetilde{U}_{N,N}^{(+)} &\widetilde{\overline{e}}_N\\
\widetilde{e}_1&\cdots&\widetilde{e}_N& g
\end{array}
\right|(x,t;y,t')\, .
\end{align*}
Expanding this result on the last column we obtain our final result
\begin{align}
\langle\boldsymbol{\psi}_{N}(\boldsymbol{q})|\Psi(x,t)\Psi^\dagger(y,t')|\boldsymbol{\psi}_{N}(\boldsymbol{q})\rangle&=
e^{i\mu(t-t')}\left[g+\frac{\6}{\6 z}\right]\det_N\left.\left(\widetilde{U}^{(+)}-z \widetilde{R}^{(+)}\right)\right|_{z=0}\, ,\\
&=e^{i\mu(t-t')}\left[\det_N\left(\widetilde{U}^{(+)}-\widetilde{R}^{(+)}\right)+(g-1)\det_N\widetilde{U}^{(+)}\right]\, ,
\end{align}
with $\widetilde{R}^{(+)}$ a $\boldsymbol{q}$-dependent matrix with elements
\be\label{defrp}
\widetilde{R}^{(+)}_{a,b}(x,t;y,t')=\widetilde{\overline{e}}_a(x,t;y,t')\widetilde{e}_b(x,t;y,t')\, , \qquad a,b=1,\cdots, N \, .
\ee

\subsection{Thermodynamic limit}\label{s24}

In order to compute the thermodynamic limit we are going to use  von Koch's determinant formula which states that for any
square matrix of dimension $M$ (which can also be infinite) denoted by $A$ and $z$ a bounded complex parameter we have \cite{Ko92}:
\begin{align}\label{vonK}
\det(1+z A)&=1+z\sum_{m=1}^M A_{m,m}+\frac{z^2}{2!}\sum_{m=1}^M\sum_{n=1}^M
\left|
\begin{array}{cc}
A_{m,m}& A_{m,n}\\
A_{n,m}& A_{n,n}
\end{array}
\right|+\cdots\, .
\end{align}

We start with $g^{(-)}(x,t;y,t')$ defined in (\ref{gm}). As a first step we need to compute the partition function $\mathcal{Z}=\mbox{Tr} \left[e^{-\mathcal{H}_I/T}\right]$.
It is easy to se that [$E_{N}(\boldsymbol{k})=\sum_{i=1}^N(\varepsilon(k_i)-\mu)$]
\begin{align}\label{zz}
\mathcal{Z}&=\sum_{N=0}^\infty\, \sum_{k_1<\cdots<k_{N}}e^{-E_{N}(\boldsymbol{k})/T}=
\prod_{k=1}^\infty\left(1+e^{-(\varepsilon(k)-\mu)/T}\right)\, .
\end{align}
Using the determinant representation for the mean value of the bilocal operator (\ref{bilocalm}) the numerator of (\ref{gm}) can be written as
\begin{align}
N^{(-)}&=\sum_{N=0}^\infty\, \sum_{k_1<\cdots<k_{N+1}}e^{-E_{N+1}(\boldsymbol{k})/T} \langle\boldsymbol{\psi}_{N+1}(\boldsymbol{k})|
\Psi^\dagger(x,t)\Psi(y,t')|\boldsymbol{\psi}_{N+1}(\boldsymbol{k})\rangle\, ,\nonumber\\
&=e^{-i\mu(t-t')}\sum_{N=0}^\infty\sum_{k_1<\cdots<k_{N+1}}e^{-E_{N+1}(\boldsymbol{k})/T}\left[\det_{N+1}\left(\widetilde{U}^{(-)}+\widetilde{R}^{(-)}\right)-\det_{N+1}\widetilde{U}^{(-)}\right]\, .
\end{align}
The energy term $e^{-E_{N+1}(\boldsymbol{k})/T}=e^{-\sum_{i=1}^{N+1}(\varepsilon(k_i)-\mu)/T}$ is distributed inside the determinants as follows:
the $i$-th row is multiplied by $e^{-(\varepsilon(k_i)-\mu)/2T}$ while the $j$-th column is  multiplied by $e^{-(\varepsilon(k_j)-\mu)/2T}$. Using
von Koch's determinant formula (\ref{vonK}) we obtain for the numerator
\be\label{i5}
N^{(-)}=e^{-i\mu(t-t')}\left[\det\left(1+\textsf{U}^{(-)}_M+\textsf{R}^{(-)}_M\right)-\det\left(1+\textsf{U}^{(-)}_M\right)\right]\, ,
\ee
with infinite dimensional determinants and  matrices $\textsf{U}^{(-)}_M$ and $\textsf{R}^{(-)}_M$  defined by
\begin{align}
[\textsf{U}^{(-)}_M]_{a,b}(x,t;y,t')&=e^{-(\varepsilon(a)-\mu)/2T}\, \textsf{U}^{(-)}_{a,b}(x,t;y,t')\, e^{-(\varepsilon(b)-\mu)/2T}\, ,\ \ a,b=1,2,\cdots\ ,\\
[\textsf{R}^{(-)}_M]_{a,b}(x,t;y,t')&=e^{-(\varepsilon(a)-\mu)/2T}\, \textsf{R}^{(-)}_{a,b}(x,t;y,t')\, e^{-(\varepsilon(b)-\mu)/2T}\, ,\ \ a,b=1,2,\cdots\ ,
\end{align}
and
\begin{align}
\textsf{U}_{a,b}^{(-)}(x,t;y,t')&=\sum_{q=1}^\infty\overline{f}(a,q|\kappa,x,t)f(b,q|\kappa,y,t')\, ,\ \ a,b=1,2,\cdots\, ,\label{defuminf}\\
\textsf{R}_{a,b}^{(-)}(x,t;y,t')&=\overline{\phi}_{a}(x,t)\phi_{b}(y,t')\, ,\ \ a,b=1,2,\cdots\, .\label{defrminf}
\end{align}
It is important to note that while $\widetilde {U}_{a,b}^{(-)}(x,t;y,t')$, $\widetilde {R}_{a,b}^{(-)}(x,t;y,t')$
were finite size matrices that were dependent on particular eigenstates $\boldsymbol{k}=(k_1,\cdots,k_{N+1})$  the matrices $\textsf{U}_{a,b}^{(-)}(x,t;y,t')$
and $ \textsf{R}_{a,b}^{(-)}(x,t;y,t')$ are infinite dimensional. From the numerator (\ref{i5}) we can extract $(1+e^{-(\varepsilon(i)-\mu)/T})^{1/2}$ from the $i$-th
row and $(1+e^{-(\varepsilon(j)-\mu)/T})^{1/2}$ from the $j$-th column obtaining an infinite product equal to  the partition function (\ref{zz}). In this way we obtain the
final result for the $g^{(-)}$ correlator in the thermodynamic limit
\be\label{gminusfinalc}
g^{(-)}(x,t;y,t')=e^{-i\mu(t-t')}\left[\det\left(1+\textsf{V}^{(T,-)}+\textsf{R}^{(T,-)}\right)-\det\left(1+\textsf{V}^{(T,-)}\right)\right]\, ,
\ee
with
\begin{align}
\textsf{V}_{a,b}^{(T,-)}(x,t;y,t')&=\sqrt{\theta(a)}\, (\textsf{U}_{a,b}^{(-)}(x,t;y,t')-\delta_{a,b}) \,  \sqrt{\theta(b)}\, ,\ \ a,b=1,2,\cdots\, ,\label{defvtminf}\\
\textsf{R}_{a,b}^{(T,-)}(x,t;y,t')&=\sqrt{\theta(a)}\, \textsf{R}_{a,b}^{(-)}(x,t;y,t')\, \sqrt{\theta(b)}\, ,\ \ a,b=1,2,\cdots\, ,\label{defrtminf}
\end{align}
where $\theta(a)=1/(1+e^{(\varepsilon(a)-\mu)/T})$ is the Fermi function.
In a similar fashion we can obtain the thermodynamic limit for the second correlation function (\ref{gp})   obtaining
\be\label{gplusfinalc}
g^{(+)}(x,t;y,t')=e^{i\mu(t-t')}\left[\det\left(1+\textsf{V}^{(T,+)}-\textsf{R}^{(T,+)}\right)+(g-1)\det\left(1+\textsf{V}^{(T,+)}\right)\right]\, ,
\ee
with $g=g(x,t;y,t')$ defined in (\ref{defg}) and
\begin{align}
\textsf{V}_{a,b}^{(T,+)}(x,t;y,t')&=\sqrt{\theta(a)}\, (\textsf{U}_{a,b}^{(+)}(x,t;y,t')-\delta_{a,b}) \,  \sqrt{\theta(b)}\, ,\ \ a,b=1,2,\cdots\, ,\label{defvtpinf}\\
\textsf{R}_{a,b}^{(T,+)}(x,t;y,t')&=\sqrt{\theta(a)}\, \textsf{R}_{a,b}^{(+)}(x,t;y,t')\, \sqrt{\theta(b)}\, ,\ \ a,b=1,2,\cdots\, .\label{defrtpinf}
\end{align}
where $\textsf{U}^{(+)}(x,t;y,t')$ and $\textsf{R}^{(+)}(x,t;y,t')$ are infinite matrices with elements
\begin{align}
\textsf{U}_{a,b}^{(+)}(x,t;y,t')&=\sum_{k=1}^\infty f(k,b|\kappa,x,t)\overline{f}(k,b|\kappa,y,t')\, ,\ \ a,b=1,2,\cdots\\
\textsf{R}^{(+)}_{a,b}(x,t;y,t')&=\overline{e}_a(x,t;y,t')e_b(x,t;y,t') \ \ a,b=1,2,\cdots\, ,
\end{align}
and $e(x,t;y,t')$,  $\overline{e}(x,t;y,t')$ are infinite vectors with elements
\begin{align}
e_a(x,t;y,t')&=\sum_{k=1}^\infty f(k,a|\kappa,x,t)\overline{\phi}_k(y,t')\, ,\ \ a=1,2,\cdots\\
\overline{e}_a(x,t;y,t')&=\sum_{k=1}^\infty \overline{f}(k,a|\kappa,y,t')\phi_k(x,t)\, ,\ \ a=1,2,\cdots\, .
\end{align}

\section{Hard-core anyons  on the lattice}\label{s3}

All the results for the correlation functions of the TG gas in the continuum can be easily extended in the case of lattice systems.
The lattice analogue of the anyonic TG gas is represented by hard-core anyons in the presence of an external potential with Hamiltonian \cite{HZC09,KLMR11}
\be\label{haml}
\mathcal{H}=-J\sum_{j=1}^{L-1}\left(a_j^\dagger a_{j+1}+a_{j+1}^\dagger a_j\right)+\sum_{j=1}^{L-1}\left(V(j,t)-\mu\right)a_j^\dagger a_j\, ,
\ee
where $J$ is the hopping parameter, $L$ is the number of lattice sites (we consider the lattice spacing $a_0=1$) and $V(j,t)$ is the confining potential.
As in the continuum case we will consider both time-independent external potentials like $V(j)=V_{0}|\left[j-(L+1)/2\right]|^\alpha$ with $\alpha=1,2,\cdots$
and time-dependent potentials implementing quantum quenches with $V(j,t=0)=V_I(j)$ and $V(j,t>0)=V_F(j)$. An example would be the sudden change of depth of
a harmonic potential characterized by $V_I(j)=V_{0}\left[j-(L+1)/2\right]^2$ and $V_F(j)=V_{1}\left[j-(L+1)/2\right]^2$ with $V_{0}\ne V_{1}$. The initial and
final Hamiltonian which governs the dynamics will be denoted by $\mathcal{H}_I$ and $\mathcal{H}_F$, respectively.
The anyonic creation and annihilation operators on different lattice sites satisfy the commutation relations
\begin{subequations}\label{comml}
\begin{align}
a_i a_j^\dagger&=-e^{- i \pi \kappa\, \mbox{\small{sign}}(i-j)}a_j^\dagger a_i+\delta_{ij}\, ,\\
a_i a_j&=-e^{i \pi \kappa\, \mbox{\small{sign}}(i-j)}a_j a_i\, ,
\end{align}
\end{subequations}
and the hard-core condition $\left\{a_j,a_j^\dagger\right\}=1\, , (a_j^\dagger)^2=\left(a_j\right)^2=0$ at the same lattice site. The commutation relations (\ref{comml})
are bosonic when $\kappa=1$  and fermionic when $\kappa=0$. In terms of fermionic operators described by  $\left\{f_i,f_j^\dagger\right\}=\delta_{ij}\, ,\left\{f_j,f_j\right\}=0$
the anyonic operators can be expressed via the generalized Jordan-Wigner transformation \cite{HZC09,KLMR11} as
\be\label{JW}
a_j^\dagger= f_j^\dagger\left(\prod_{\beta=1}^{j-1}e^{- i\pi\kappa f_\beta^\dagger f_\beta} \right)\, ,
\ \ a_j= \left(\prod_{\beta=1}^{j-1}e^{i\pi\kappa f_\beta^\dagger f_\beta} \right)f_j\, ,
\ee
which shows that the hard-core anyonic operators are products of an odd number of fermionic operators. We consider open boundary conditions for finite size systems.
In the bosonic case the Hamiltonian (\ref{haml})
is equivalent with the Hamiltonian of the XX spin-chain (\cite{LSM61}, Chap.~I of \cite{FranB}) after the identification $a_j=\sigma_j^+$, $a_j^\dagger=
\sigma_j^-$.   We should point out certain characteristics of impenetrable particles systems on the lattice which makes their study   extremely worthwhile.
While at low filling fractions the results for lattice systems reproduce the ones for continuum systems at moderate and large fillings the presence of the
lattice introduces new physics like  the Mott insulator phase \cite{BRSM02}  and the emergence of quasicondensates  at finite momentum after
expansion \cite{RM04b,VRSB15}.  From the computational point of view finite size systems on the lattice have the advantage of being easier to access numerically due to the
finite size of the  associated single particle Hilbert space.  Determinant representations for some correlators (static or equal time at nonequilibrium) and
the  dynamics of hard-core particles on the lattice in  various nonequilibrium scenarios  have been previously  investigated in
\cite{BMD70,BM71a,BM71b,MBA71,ARRS99,RM04b,RM04a,Rig05,RM05,RM06,RMO06,PK07,RDYO07,HZC09,HC12,VXR17,Hao16,XR17,LSP19,GLC20,SKPD21,ASW22,GDE22}.

Introducing the Fock vacuum defined by $a_j|0\rangle=\langle 0|a_j^\dagger=0$ for all $j$ the eigenstates of the system at $t=0$ are
$\boldsymbol{k}=(k_1,\cdots,k_N)$ \cite{HZC09, KLMR11, Pat15}:
\begin{align}\label{eigenl}
|\boldsymbol{\psi}_N(\boldsymbol{k}\rangle=&\frac{1}{\sqrt{N!}}\sum_{m_1=1}^{L-1} \cdots\sum_{m_N=1}^{L-1} \,  \psi_{N}(m_1,\cdots,m_N|\boldsymbol{k})\,
a_{m_N}^\dagger\cdots a_{m_1}^\dagger|0\rangle\, ,
\end{align}
with the many-body anyonic wave function given by
\be\label{wavefl}
\psi_{N}(m_1,\cdots,m_N|\boldsymbol{k})=\frac{i^{N(N-1)/2}}{\sqrt{N!}}\prod_{1\le a<b\le N}e^{i\frac{\pi \kappa}{2}\epsilon(m_a-m_b)}
\det\left[\phi_{k_a}(m_b)\right]_{a,b=1,\cdots,N}\, ,
\ee
and $\epsilon(m)=1$ for $m\ge 0$ and $\epsilon(m)=-1$ for $m<0$. We should point out that the value of $\epsilon(0)$ is not physically relevant
(the wave function (\ref{wavefl}) is zero when two coordinates are equal, see also the discussion in \cite{CIKT93}) but our previous choice makes
some numerical computations easier. In (\ref{wavefl}) $\phi_k(m)$ are the single particle fermionic wave functions  satisfying
$\mathcal{H}_I^{\tiny{SP}}(m)\phi_k(m)=\varepsilon(k)\phi_k(m)$  with  $\mathcal{H}_I^{\tiny{SP}}=-J\sum_{m=1}^{L-1}(|m+1\rangle \langle m|+|m\rangle\langle m+1|)
+\sum_{m=1}^{L-1}(V(m,t=0)-\mu)|m\rangle\langle m|\, $ and $|m\rangle=f_m^\dagger | 0\rangle$.
The eigenstates (\ref{eigenl}) form a complete set, are normalized  and satisfy $\mathcal{H}_I |\boldsymbol{\psi}_N(\boldsymbol{k})\rangle=
E_N(\boldsymbol{k})|\boldsymbol{\psi}_N\rangle $  with $E_N(\boldsymbol{k})=\sum_{i=1}^N(\varepsilon(k_i)-\mu)\, .$

The dynamics of the eigenstates is given by  $|\boldsymbol{\psi}_N(\boldsymbol{k,t})\rangle=e^{- i t \mathcal{H}_F} |\boldsymbol{\psi}_N(\boldsymbol{k})\rangle$
with the time-dependent many-body wavefunction
\be\label{waveflt}
\psi_{N}(m_1,\cdots,m_N|\boldsymbol{k},t)=\frac{i^{N(N-1)/2}}{\sqrt{N!}}\prod_{1\le a<b\le N}e^{i\frac{\pi \kappa}{2}\epsilon(m_a-m_b)}
\det\left[\phi_{k_a}(m_b,t)\right]_{a,b=1,\cdots,N}\, ,
\ee
where in the case of a time-independent external potential ($\mathcal{H}_I=\mathcal{H}_F$) we have  $\phi_k(m,t)=e^{-i \varepsilon(k) t}\phi_k(m)$
 while in the case of a quantum quench ($\mathcal{H}_I\ne\mathcal{H}_F$)  $\phi_k(m,t)$ is the unique time-dependent solution of the Schr\"odinger equation
$i\hbar \6 \phi_{k}(m,t)/\6 t=H_F^{\mbox{\tiny{SP}}}(m) \phi_{k}(m,t)$ with $H_F^{\mbox{\tiny{SP}}}(m)=-J\sum_{m=1}^{L-1}(|m+1\rangle \langle m|+|m\rangle\langle m+1|)
+\sum_{m=1}^{L-1}(V(m,t>0)-\mu)|m\rangle\langle m|\,$ and initial condition $\phi_k(m,0)=\phi_k(m)$. As in the continuum case our formalism is not valid for
 systems with $V(j,t)=0$ and  periodic boundary conditions. For this situation the Fredholm determinant representations for the correlators can be found in \cite{CIKT93,Pat15}.


\section{Determinant representations for finite temperature correlators on the lattice }\label{s4}

The determinant representations for the field correlators can be derived in a similar fashion as in the continuum case.
The relevant correlation functions are now:
\begin{align}
g^{(-)}(x,t;y,t')&=\langle a^\dagger_x(t)a_y(t')\rangle_{\mu,T}
=\mbox{Tr} \left[e^{-\mathcal{H}_I/T}a^\dagger_x(t)a_y(t')\right]/\mbox{Tr} \left[e^{-\mathcal{H}_I/T}\right]\, ,\label{gml}\\
g^{(+)}(x,t;y,t')&=\langle a_x(t) a_y^\dagger (t')\rangle_{\mu,T}
=\mbox{Tr} \left[e^{-\mathcal{H}_I/T}a_x(t) a_y^\dagger (t')\right]/\mbox{Tr} \left[e^{-\mathcal{H}_I/T}\right]\, ,\label{gpl}
\end{align}
where $a^\dagger_x(t)=e^{ i \mathcal{H}_F t}a^\dagger_x e^{- i \mathcal{H}_F t}$ and $a_x(t)=e^{ i \mathcal{H}_F t}a_x e^{- i \mathcal{H}_F t}$ .
The spectral function is defined as
\be\label{spectrall}
A(k,\omega)=-\frac{1}{\pi}\mbox{Im } G^R(k,\omega)\, ,\ \ \ G^R(k,\omega)=\inti dt\,  e^{i\omega t}\sum_{x,y}e^{- ik(x-y)}G^R(x,t;y,0)\, ,
\ee
with the retarded Green's function in real space and time defined by
\begin{align}\label{spectralfl}
G^R(x,t;y,t')&=-i \Theta(t-t')\left[\langle a_x(t)a_y^\dagger(t')\rangle_{\mu,T}+\langle a_y^\dagger(t')a_x(t)\rangle_{\mu,T}\right]\, ,\nonumber\\
&=-i \Theta(t-t')\left[g^{(+)}(x,t;y,t')+g^{(-)}(y,t';x,t)\right]\, .
\end{align}
We need to clarify an important point. In general the retarded Green's function for bosonic operators is defined as the difference between the
greater and lesser Green's functions and, therefore, one might expect a minus sign on the right hand side of (\ref{spectralfc}) and (\ref{spectralfl}).
However as it can be seen from the Jordan-Wigner transformation (\ref{JW})  the anyonic operators can be expressed in terms of a product of an odd number
of fermionic operators and in this case the spectral function is defined as the sum of the greater and lesser Green's functions as it is discussed in
Chap. 3.3. of \cite{Mah00}.
 Using $\langle a_x(t)a_y^\dagger(0)\rangle_{\mu,T}=\overline{\langle a_x(-t)a_y^\dagger(0)\rangle_{\mu,T}}$  we obtain
\be
A(k,\omega)=\inti dt\,  e^{i\omega t}\sum_{x,y}e^{- ik(x-y)}
\left[g^{(+)}(x,t;y,0)+g^{(-)}(y,0;x,t)\right]\, .
\ee

The determinant representations for $g^{(\pm)}(x,t;y,t')$ on the lattice are given by the same expressions as in the continuum case,
(\ref{gminusfinalc}) and (\ref{gplusfinalc}), the only difference being that in the definitions of the $\textsf{V}^{(T,\pm)}(x,t;y,t')$
and $\textsf{R}^{(T,\pm)}(x,t;y,t')$ kernels defined in (\ref{defvtminf}), (\ref{defvtpinf}), (\ref{defrtminf}) and (\ref{defrtpinf}) now
the $f(k,q|\kappa,x,t)$ function is given by
\be\label{deffl}
f(k,q|\kappa,x,t)=\delta_{k,q}-(1-e^{i\pi\kappa})\sum_{x'=x}^{\infty} \phi_k(x',t)\overline{\phi}_q(x', t) \, .
\ee

\section{Comparison with previous results}\label{s5}

For an easier numerical implementation, but also in order to compare with previously derived similar results,
it is useful to express the representations derived in Sect.~\ref{s24} in terms of products and determinants
of simple matrices. First we introduce a column vector defined by $\boldsymbol{\phi}(x,t)=\left(\phi_1(x,t),\cdots,\phi_M(x,t)\right)^{T}$
where $M$ is the dimension of the single particle Hilbert space (or the truncated dimension). For continuum models $M$ is infinite
but in the case of finite size lattice systems $M$ is finite. The adjoint of $\boldsymbol{\phi}(x,t)$ is $\boldsymbol{\phi}^\dagger(x,t)=
\overline{\boldsymbol{\phi}}^T(x,t)=\left(\overline{\phi}_1(x,t),\cdots,\overline{\phi}_M(x,t)\right)$. We also need to introduce two
additional  matrices
\be
\boldsymbol{P}_{a,b}(x,t|\kappa)=f(a,b|\kappa,x,t)\, ,\ \ \ \boldsymbol{F}_{a,b}=\delta_{a,b}\sqrt{\theta(a)}\, , \ \  a,b,=1,2,\cdots ,
\ee
with $f(a,b|\kappa,x,t)$ defined in (\ref{deff}) for continuum systems and in (\ref{deffl}) for lattice systems and $\theta(a)=1/(1+e^{(\varepsilon(a)-\mu)/T})$ is the Fermi function.
 Using $\overline{f}(a,b|\kappa,x,t)=f(b,a|-\kappa,x,t)$ it can be shown that $\textsf{V}^{(T,\pm)}$ and $\textsf{R}^{(T,\pm)}$ defined in
(\ref{defvtminf}), (\ref{defvtpinf}), (\ref{defrtminf}) and (\ref{defrtpinf}) can be expressed in terms of the previous matrices as follows
($\boldsymbol{1}$ is the identity matrix) :
\begin{subequations}\label{matrixf}
\begin{align}
\textsf{V}^{(T,-)}(x,t;y,t')&=\boldsymbol{F}\left[\left(\boldsymbol{P}(y,t'|\kappa)\boldsymbol{P}(x,t|-\kappa)\right)^T-\boldsymbol{1}\right]\boldsymbol{F}\, =
\left[\boldsymbol{F}\left(\boldsymbol{P}(y,t'|\kappa)\boldsymbol{P}(x,t|-\kappa)-\boldsymbol{1}\right)\boldsymbol{F}\right]^T\, ,\\
\textsf{V}^{(T,+)}(x,t;y,t')&=\boldsymbol{F}\left[\boldsymbol{P}(y,t'|-\kappa)\boldsymbol{P}(x,t|\kappa)-\boldsymbol{1}\right]\boldsymbol{F}\, ,\\
\textsf{R}^{(T,-)}(x,t;y,t')&=\boldsymbol{F}\, \overline{\boldsymbol{\phi}}(x,t) \boldsymbol{\phi}^T(y,t') \boldsymbol{F}\, ,\\
\textsf{R}^{(T,+)}(x,t;y,t')&=\boldsymbol{F}\,\left[\boldsymbol{P}(y,t'|-\kappa)\boldsymbol{\phi}(x,t)\right] \left[\boldsymbol{\phi}^\dagger(y,t')\boldsymbol{P}(x,t|\kappa)\right]  \boldsymbol{F}
\end{align}
\end{subequations}
and also $g(x,t;y,t')=\boldsymbol{\phi}^\dagger(y,t')\boldsymbol{\phi}(x,t)$. Our results for the correlation functions (\ref{gminusfinalc}), (\ref{gplusfinalc})
can be rewritten using a simple formula which states that for any square matrix of dimension $M$ and $u$ and $v$ two column vectors of the same dimension
the following relation is valid \cite{Marc90}
\begin{align}
\det(A+uv^T)=\det A+\det A\, v^TA^{-1}u\, .
\end{align}
Using this formula and (\ref{matrixf})  we find
\begin{subequations}\label{resmatrixc}
\begin{align}
g^{(-)}(x,t;y,t')&=e^{-i\mu(t-t')}\left(\det\boldsymbol{W}_{(-)}\right)\boldsymbol{\phi}^T(y,t')\boldsymbol{F}\,\boldsymbol{W}_{(-)}^{-1}\,\boldsymbol{F}\overline{\boldsymbol{\phi}}(x,t)\, ,\\
g^{(+)}(x,t;y,t')&=e^{i\mu(t-t')}\left(\det\boldsymbol{W}_{(+)}\right)\left[\boldsymbol{\phi}^\dagger(y,t')\boldsymbol{\phi}(x,t)-
\boldsymbol{\phi}^\dagger(y,t')\boldsymbol{P}(x,t|\kappa)\boldsymbol{F}\, \boldsymbol{W}_{(+)}^{-1}\, \boldsymbol{F}\boldsymbol{P}(y,t'|-\kappa)\boldsymbol{\phi}(x,t)\right]\, ,
\end{align}
\end{subequations}
with $\boldsymbol{W}_{(-)}=\boldsymbol{1}+\left[\boldsymbol{F}\left(\boldsymbol{P}(y,t'|\kappa)\boldsymbol{P}(x,t|-\kappa)-\boldsymbol{1}\right)\boldsymbol{F}\right]^T$
and $\boldsymbol{W}_{(+)}=\boldsymbol{1}+\boldsymbol{F}\left[\boldsymbol{P}(y,t'|-\kappa)\boldsymbol{P}(x,t|\kappa)-\boldsymbol{1}\right]\boldsymbol{F}$. In the case
of time-independent potentials similar results were derived by Wang in \cite{W22}.
The final results (\ref{resmatrixc}) are expressed as sums, products and determinants of matrices with elements given by
partial overlaps of the single particle wavefunctions, (\ref{deff}) and (\ref{deffl}). In many experimentally relevant situations these overlaps can be calculated
analytically like in the case of continuum systems with harmonic trapping \cite{AGBK17}, triangular potential \cite{VS},  Dirichlet and Neumann boundary conditions
or very easily numerically by quadratures or simple summations (this of course requires as a preliminary step the determination of the dynamically evolved single
particle eigenfunctions  which are obtained by solving the appropriate  time-dependent Schr\"odinger equation). This makes our results  (\ref{resmatrixc}) extremely
easy and efficient to implement numerically requiring three simple steps: (i) computation of the time-evolved single-particle eigenfunctions; (ii) calculation of the $\boldsymbol{P}$
and $\boldsymbol{F}$ matrices and $\boldsymbol{\phi}(x,t)$,  $\boldsymbol{\phi}(y,t')$ vectors; (iii) determination of the $g^{(\pm)}(x,t;y,t')$ correlators using (\ref{resmatrixc}).
Let us look at some particular cases where results for the correlators are known.

\textit{Free fermions.} In the case of free fermions, $\kappa=0$, we have $f(a,b|0,x,t)=\delta_{a,b}$ and, therefore, the matrix $\boldsymbol{P}$ reduces to the
identity matrix. We obtain
\begin{align}
g^{(-)}(x,t;y,t')&=e^{-i\mu(t-t')}\sum_{a}\theta(a)\overline{\phi}_a(x,t)\phi_a(y,t')\, ,\\
g^{(+)}(x,t;y,t')&=e^{i\mu(t-t')}\sum_{a}(1-\theta(a))\overline{\phi}_a(y,t')\phi_a(x,t)\, ,
\end{align}
which in the equilibrium case, $\phi_a(x,t)=e^{-i \varepsilon(a) t}\phi_a(x)$, are the well known results for free fermions at finite temperature.

\textit{Zero temperature.} At zero temperature the elements of the $\boldsymbol{F}$  matrix  become $\boldsymbol{F}_{a,b}=\delta_{a,b}$ for $a\le N$
and $\boldsymbol{F}_{a,b}=0$ for $a>N$ where $N$ is the number of particles in the ground state. The multiplication with $\boldsymbol{F}$   acts as an
projector on the first $N$ states obtaining $\boldsymbol{\phi}^T(y,t')\boldsymbol{F}=\left(\phi_1(y,t'),\cdots,\phi_N(y,t'),0,0,\cdots\right)^{T}$,
$\boldsymbol{F}\overline{\boldsymbol{\phi}}(x,t)=\left(\phi_1(x,t),\cdots,\phi_N(x,t),0,0,\cdots\right)$ and so on.
For the relevant matrices we find  $[\boldsymbol{W}_{(-)}]_{ab}=\left[\boldsymbol{P}(y,t'|\kappa)\boldsymbol{P}(x,t|-\kappa)\right]^T_{ab}$ for $a,b\le N$  and $[\boldsymbol{W}_{(-)}]_{ab}=\delta_{ab}$
for $a> N$ or $b>N$ and $[\boldsymbol{W}_{(+)}]_{ab}=\left[\boldsymbol{P}(y,t'|\kappa)\boldsymbol{P}(x,t|-\kappa)\right]_{ab}$ for $a,b\le N$  and $[\boldsymbol{W}_{(+)}]_{ab}=\delta_{ab}$
for $a> N$ or $b>N$. Therefore, at zero temperature we obtain
\begin{align}
g^{(-)}(x,t;y,t')&=e^{-i\mu(t-t')}\det\left[\boldsymbol{P}(y,t'|\kappa)\boldsymbol{P}(x,t|-\kappa)\right]\times\boldsymbol{\phi}^T(y,t')\left[\boldsymbol{P}(y,t'|\kappa)\boldsymbol{P}(x,t|-\kappa)\right]^{-1 T}
\overline{\boldsymbol{\phi}}(x,t)\, ,\\
g^{(+)}(x,t;y,t')&=e^{i\mu(t-t')}\det \left[\boldsymbol{P}(y,t'|-\kappa)\boldsymbol{P}(x,t|\kappa)\right]\nonumber\\
&\qquad \times\left[\boldsymbol{\phi}^\dagger(y,t')\boldsymbol{\phi}(x,t)-
\boldsymbol{\phi}^\dagger(y,t')\boldsymbol{P}(x,t|\kappa)\,\left[\boldsymbol{P}(y,t'|-\kappa)\boldsymbol{P}(x,t|\kappa)\right]^{-1}
\boldsymbol{P}(y,t'|-\kappa)\boldsymbol{\phi}(x,t)\right]\, ,
\end{align}
where in the above expressions with the exception of $\boldsymbol{\phi}^\dagger(y,t')\boldsymbol{\phi}(x,t)$ all the determinants and products of matrices
need to be understood as projections on the first $N$ states.  In the case of impenetrable bosons ($\kappa=1$)  this is the result obtained by Settino
\textit{et. al.} in \cite{SGPM21} (in our result we have  extracted the dependence on the chemical potential in the first term).

\textit{Equal time case of $g^{(-)}(x,t;y,t')$. } The simplifications that appear in the  equal time case of the $g^{(-)}$ correlator  are investigated in Appendix \ref{a1}. In this case
the determinant representation in the continuum case is considerable simpler
\be\label{gmsmall}
g^{(-)}(x,t;y,t)=\det\left(1+\textsf{v}^{(T,-)}+\textsf{r}^{(T,-)}\right)-\det\left(1+\textsf{v}^{(T,-)}\right)\, ,
\ee
with
\begin{align}
\textsf{v}_{a,b}^{(T,-)}(x,t;y,t)&=\,-(1-e^{-i\pi\kappa\mbox{\small{sign}}(y-x)})\mbox{sign}(y-x)\sqrt{\theta(a)\theta(b)} \int_{x}^{y}\overline{\phi}_a(v,t)\phi_b(v,t)\, dv \, \, ,\ \ a,b=1,2,\cdots\, ,\label{vsmall}\\
\textsf{r}_{a,b}^{(T,-)}(x,t;y,t)&=\sqrt{\theta(a)}\, \overline{\phi}_a(x,t)\phi_b(y,t)\, \sqrt{\theta(b)}\, ,\ \ a,b=1,2,\cdots\, .\label{rsmall}
\end{align}
In the lattice case (\ref{gmsmall}) remains valid the only difference is that the $\textsf{v}_{a,b}^{(T,-)}$ kernel is now
\begin{align}
\textsf{v}_{a,b}^{(T,-)}(x,t;y,t)&=\,-(1-e^{-i\pi\kappa\mbox{\small{sign}}(y-x)})\sqrt{\theta(a)\theta(b)}\sum_{m=\min(x,y)}^{\max(x,y)-1}\overline{\phi}_a(m,t)\phi_b(m,t) \, \, ,\ \ a,b=1,2,\cdots\, .\label{vsmalll}
\end{align}
For the continuous bosonic system $(\kappa=1)$ the representations (\ref{gmsmall}) at zero and finite temperature were derived in \cite{PB07}  and
\cite{AGBK17}, respectively, while  the anyonic generalization was obtained in \cite{P20} (please note that in \cite{AGBK17,P20} the
$g^{(-)}(x,t;y,t)$ correlator is related to the one presented in this paper by complex conjugation). In the case of  lattice systems the
representation (\ref{gmsmall}) with (\ref{vsmalll}), which represents the lattice generalization of Lenard's formula \cite{Len66}, to our knowledge
has not been reported previously in the literature except for the particular case of  bosons at zero temperature in equilibrium  \cite{NI11}.

\section{Spectral function for a trapped Tonks-Girardeau lattice gas}\label{s6}

\begin{figure*}
\includegraphics[width=1\linewidth]{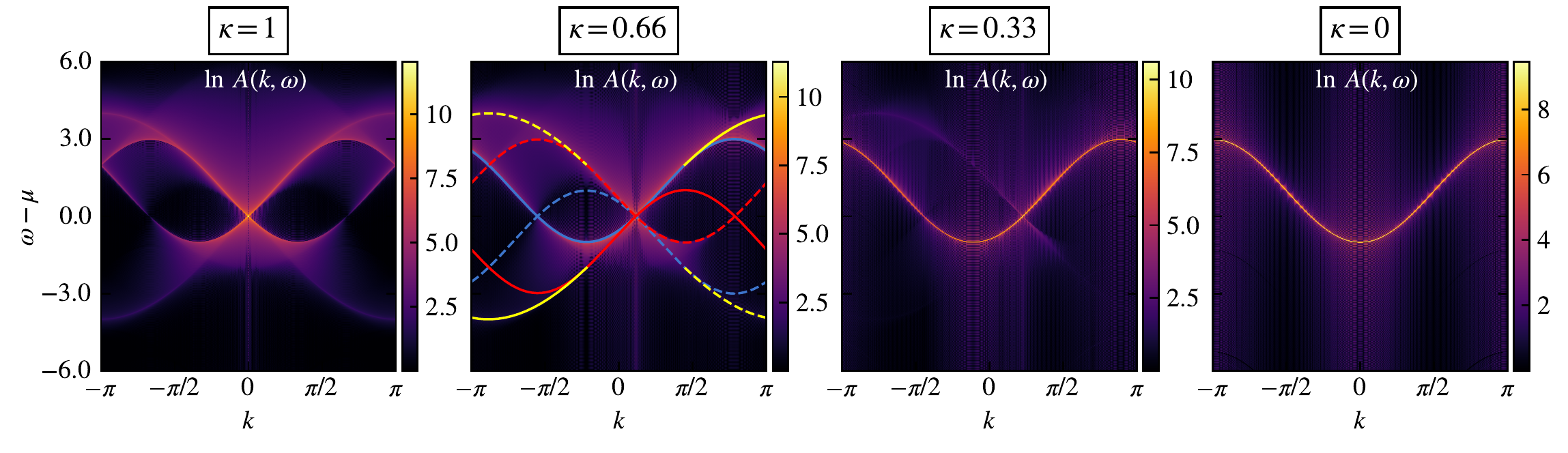}
\caption{Logarithm of the zero temperature spectral function $\ln A(k,\omega)$ of a TG gas on the lattice for different values
of the statistics parameter. The relevant parameters are $L=300$, $N=100$ and  $V(j,t\ge 0)=0$. In the second panel the continuous blue, red and
yellow lines mark the $\varepsilon_I(k)$ , $\varepsilon_{II}(k)$ and $\varepsilon_{III}(k)$ singular lines [see Eqs.~(\ref{e1}), (\ref{e2}), (\ref{e3})] while the corresponding
dashed lines describe $-\varepsilon_I(k)$ , $-\varepsilon_{II}(k)$ and $-\varepsilon_{III}(k)$.
}
\label{fig1}
\end{figure*}

In this section we are going to investigate the influence of a harmonic trapping potential on the spectral function
of a TG gas on the lattice. The spectral function $A(k,\omega)$ defined in (\ref{spectrall}) quantifies the probability for exciting a particle(hole) with
energy $\omega (-\omega)$ and momentum $k$. In general this probability interpretation is not valid in the case of bosonic systems due to the fact that
for these systems $A(k,\omega)$ can be negative. However in the case of impenetrable systems of any statistics it can be shown \cite{W22} that both contributions on the
right hand side of (\ref{spectrall}) are positive and therefore the probabilistic interpretation is still valid.
In Fig.~\ref{fig1} we present the logarithm of the zero temperature spectral function $\ln A(k,\omega)$ for a finite size system with
$L=300$ and $N=100$ particles for different values of the statistics parameter and $V(j,t\ge0)=0$. There are three main singularity lines \cite{SGPM21,W22} and below we are going
to derive their explicit expressions. We will focus on the bosonic case as the results for different statistics can be obtained by a
simple momentum shift with $\kappa (k_F-1)$ (see Fig.7 of \cite{P19} and the mean field discussion in \cite{W22}). For a system of $N$ particles on a
lattice with $L$ sites with no external potential and  open boundary conditions the single particle dispersion is $\varepsilon(k)=-2J \cos(k)$ with the
Fermi vector $k_F=\pi N/L$ and  the chemical potential given by $-2J \cos k_F=\mu$. The first singular line corresponds to the case when a particle is
excited from the Fermi level $k_F$ to $k_F+q$. This particle-type excitation is also present in continuous systems (Type I excitation in Lieb's
classification \cite{L63}) and has energy $-2J \cos(k_F+q)+2 J\cos k_F$ and momentum $k=k_F+q-k_F$. Using the definition of the chemical potential
the first singularity line is $\varepsilon_I(k)=-2J \cos(k_F+k)-\mu$. The second line is due to hole-type excitations (Type II excitations in
Lieb's classification) in which a particle from the Fermi sea with quasimomentum $-k_F\le k\le k_F$ moves to the first unoccupied state above the
Fermi level i.e., $k_F+\pi/L$. Neglecting $1/L$ corrections, the energy and momentum of these excitations are $-2J\cos k_F+2J\cos q$  and $k=k_F-q$
respectively. Therefore, the singularity line is give by $\varepsilon_{II}(k)=2J\cos(k_F-k)+\mu$. The third line has no equivalent in continuous
systems and is generated by excitations from an occupied state inside the Fermi sea $q$ to a state with quasimomentum $\pi -q$. The energy and momenta
of these excitations are $-2J\cos(\pi -q)+2J\cos q$ and $k=\pi-2q$ and the line is described by $\varepsilon_{III}(k)=-2J\cos[(\pi+k)/2] +2J\cos[(\pi-k)/2]=4J\sin(k/2)$.
The general result valid for any value of the statistics parameter $\kappa$ is obtained  by shifting $k\rightarrow k+k_F(\kappa-1)$. We find
\begin{subequations}
\begin{align}
\varepsilon_{I}(k)&=-2J\cos(k+k_F\kappa)-\mu\, ,\label{e1}\\
\varepsilon_{II}(k)&=2J\cos(k+k_F\kappa-2k_F)+\mu\, ,\label{e2}\\
\varepsilon_{III}(k)&=4 J\sin[(k+k_F\kappa-k_F)/2]\, .\label{e3}
\end{align}
\end{subequations}
\begin{figure*}
\includegraphics[width=1\linewidth]{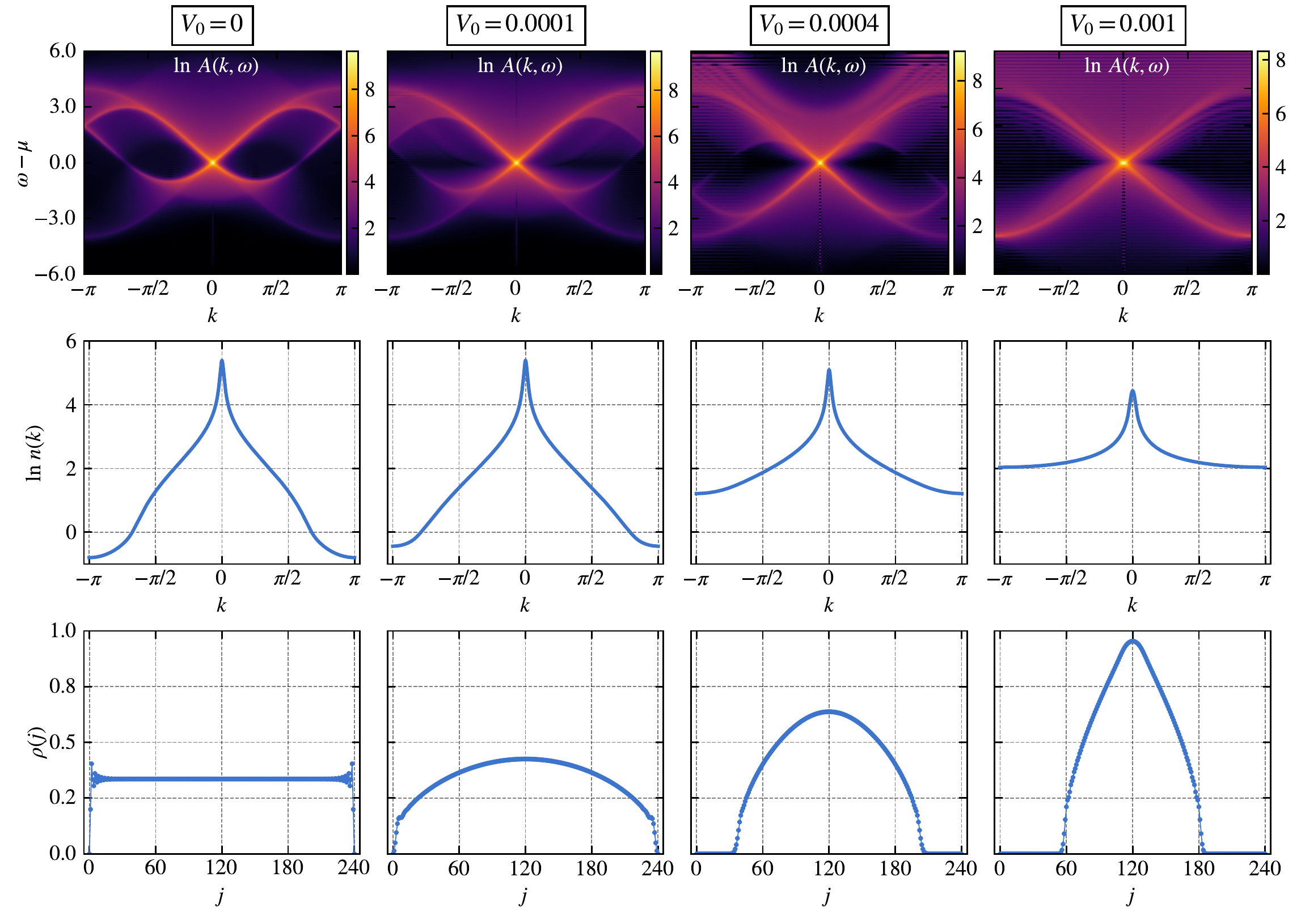}
\caption{First row: Logarithm of the spectral function $\ln A(k,\omega)$ for a harmonically trapped $(V(j,t\ge 0)=V_{0}\left[j-(L+1)/2\right]^2)$
bosonic TG gas  on the lattice at temperature $T=0.05$ and different values of the curvature $V_0$. The other relevant parameters are
$L=240$ and $N=80$. The second row present the logarithm of the momentum distribution $\ln n(k)$ and the third row the density of particles $\rho(j)$  on
the lattice.
}
\label{fig2}
\end{figure*}
In the case of homogeneous continuous systems the non-linear Luttinger theory \cite{ISG12} predicts that the spectral function has a power-law
behaviour $A(k,\omega)\sim|\omega-\varepsilon_{j}(k)|^{-\mu_j}$ with $j=\{I,II\}$ near each singular line $\varepsilon_j(k)$. The spectral
function of a lattice system at small filling fractions presents power law behaviour near the singular lines with power exponents
which are very close to the nonlinear Luttinger liquid predictions \cite{SGPM21,W22}. From Fig.~\ref{fig1} we see that as the statistics
parameter decreases the spectral weight from $\varepsilon_{II}(k)$ and $\varepsilon_{III}(k)$ also decreases being transferred to
$\varepsilon_{I}(k)$ which becomes the only singular line at $\kappa=0$  as it is expected for a free fermionic system.

Fig.~\ref{fig2} presents results for the spectral function, momentum distribution and density profile for a system of hard-core bosons $(\kappa=1)$ at temperature
$T=0.05$ in the presence of a harmonic trapping potential $V(j,t\ge 0)=V_{0}\left[j-(L+1)/2\right]^2$ of different curvatures. The presence of the
confining potential has a dramatic effect on both static and dynamic properties of the system. While in the case of open boundary conditions the
density is almost constant with small distortions along the boundaries  from the last row of Fig.~\ref{fig2} we see that as we increase the strength
of the harmonic potential the density profile becomes nonuniform with a large number of particles concentrated in the center of the trap and very
few particles at the edges. At a critical value of the curvature (the number of particles is kept constant) the density of the particles in the
center of the trap will become one and by further increasing $V_0$ this insulating region with zero compressibility will further increase. What we
have described is the well known superfluid to Mott insulator quantum phase transition (QPT)  \cite{CCGO11} induced by the variation of the potential's curvature.
The microscopic origin of this QPT has been investigated in \cite{RM04c}. While in the case of trapped continuum system
the energy level spacing of the single particle system is constant, in the  lattice case the level spacing decreases with increasing level number until
a degeneracy sets in and the level spacing becomes monotonously increasing. The  eigenfunctions of the degenerate states
have the interesting characteristic of having zero weight in the middle of the trap and as the level increases this region in which the eigenfunctions
have zero weight also increases. The QPT occurs  when the Fermi energy is close to the level where this degeneracy  sets in and the density in the
center of the trap approaches one. Further increasing the curvature of the potential  results in a similar increase of the insulating region and due
to the Pauli principle the eigenfunctions should also have an increasing region with zero weight in the center of the trap as the level increases,
which is exactly the phenomenon mentioned above.

Signatures of the QPT can also been seen in the momentum distribution which due to the presence of the lattice is a periodic function  in the reciprocal
lattice. While in the case of bosons and fermions the momentum distribution $n(k)$ is symmetric with respect to $k=0$ we should point out that for
anyonic systems this is no longer valid as a result of the broken space-reversal symmetry of the commutation relations (\ref{comml}).  At zero temperature
and small values of the curvature there is always a region in which $n(k)=0$ but as we increase $V_0$ another way of identifying the presence of the
insulating phase is the fact that $n(k=\pm \pi)$ becomes nonzero \cite{RM04c}. As the size of the insulator region increases this is also accompanied by a similar increase
of $n(k=\pm\pi)$. At finite but low  temperature we can see from the second row of Fig.~\ref{fig2} that we encounter a similar phenomenon. When $V_0=0$ the momentum distribution
is very close to the usual quasicondensate behaviour at $T=0$ with $n(k)\sim k^{-1/2}$ at $k=0$  and $n(k=\pm\pi)\sim 0$ (this is due to thermal fluctuations). As we increase the strength of the confining potential
 $n(k=\pm\pi)$ also increases and as an insulating plateau develops we see that $n(k)$ becomes almost flat with a smooth peak at  $k=0$.

As the spectral function quantifies the probability that a particle (hole) to be excited (filled) at a given momentum $k$  and energy $\omega$
and the tails of the momentum distribution become more populated with increasing curvature one would expect that  the spectral weight would get transferred
to the hole component $(\omega\le \mu)$ of $A(k,\omega)$ as $V_0$ increases. This is indeed the case and  it can be seen in the first row of Fig.~\ref{fig2}.
An interesting peculiarity of one dimensional systems is that at low energies the spectral function has a region in which it is zero (or it is very small at
finite temperatures). In our case for the homogeneous system at $T=0$ this region is defined by  $|\omega-\mu|<\varepsilon_{II}(k)$ and is due to
the fact that the Fermi sphere in 1D is comprised of only two points $\pm k_F$. While in higher dimensions the hole excitations lead to a continuum extending to zero energy for
all momenta $q$ smaller than $2 k_F$ in 1D the only place where the hole excitations can reach zero are for $q=0$ and $q=2k_F$.
At moderate values of the curvature,  $V_0=0.0001$, we see that the region in which the spectral function vanishes  for a homogeneous system decreases considerably
but as  we further increase the curvature  the opposite phenomenon occurs. An interesting feature is revealed at $V_0=0.0004$ where we can see that $A(k,\omega\le \mu)$
is almost similar with $A(k,\omega\ge \mu)$ for a homogeneous system ($V_0=0$).
For $V_0=0.001$ and zero temperature the system would develop an insulator region in the middle of the trap. In this case we see from Fig.~\ref{fig2}
that the spectral function  presents only two singular lines compared with the three in the case of the homogeneous system or for the trapped system
before the QPT. Therefore, the measurement of the spectral function would provide an alternative way of identifying the presence of the Mott insulator phase
in a trapped 1D system.

\section{Conclusions}\label{s7}

We have derived determinant representations for the Green's functions and spectral function of impenetrable particles of arbitrary
statistics on the lattice and in the  continuum. Our results are valid at zero and finite temperature and for general confining potentials
including nonequlibrium scenarios due to the sudden changes of the external potential. The main advantage of our approach is the numerical
efficiency with the main computational cost being the calculation of partial overlaps of the dynamical evolving single particle basis.
Our numerical analysis of the spectral function for a trapped TG gas on the lattice  showed that the presence of the external potential
has a profound effect on the distribution of the spectral weight
culminating with the vanishing of one of the spectral lines present in the homogeneous system as we increase the curvature of the potential.
In addition, our results also constitute the starting point in the rigorous analysis of the asymptotics of the correlation functions using
powerful techniques developed in the last decades \cite{KBI,KKMS12,GKKK17,GIZ21} and can be used for benchmarking numerical work in the
finite coupling case \cite{PC14,PS21}.
A natural generalization of our results would be the derivation of similar representations for the correlators of impenetrable multicomponent
systems like the Gaudin-Yang and Hubbard models. This will be deferred to a future publication.

\acknowledgments
Financial support  from the grant no. 16N/2019 of the National Core Program  of the Romanian Ministry of Research,
Innovation, and Digitization is gratefully acknowledged.

\appendix

\section{The equal time case of $g^{(-)}(x,t;y,t')$ }\label{a1}

In this appendix we  investigate how the $g^{(-)}(x,t;y,t')$ correlator simplifies when $t=t'$.
In the continuum case we have $\textsf{U}_{a,b}^{(-)}(x,t;y,t)=\sum_{q=1}^\infty\overline{f}(a,q|\kappa,x,t)f(b,q|\kappa,y,t)$ with $f(a,q|\kappa,x,t)$ defined in (\ref{deff}).
We introduce $\xi=(1-e^{i\pi\kappa})$ and consider first the case $x\le y$. We have
\begin{align}\label{i6}
\textsf{U}_{a,b}^{(-)}(x,t;y,t)-\delta_{a,b}&=-\overline{\xi}\int_{x}^{L_+}\overline{\phi}_a(v,t)\phi_b(v,t)\, dv- \xi\int_{y}^{L_+}\overline{\phi}_a(w,t)\phi_b(w,t)\, dw\nonumber\\
&\qquad\qquad\qquad\qquad +\xi\,\overline{\xi}\sum_{q=1}^\infty\left(\int_{x}^{L_+}\overline{\phi}_a(v,t)\phi_q(v,t)\, dv\right)\left(\int_{y}^{L_+}\phi_b(w,t)\overline{\phi}_q(w,t)\, dw\right)\, ,\nonumber\\
&= -\overline{\xi} \int_{x}^{y}\overline{\phi}_a(v,t)\phi_b(v,t)\, dv-(\xi+\overline{\xi})\int_{y}^{L_+}\overline{\phi}_a(w,t)\phi_b(w,t)\, dw\nonumber\\
&\qquad\qquad\qquad\qquad +\xi\,\overline{\xi}\sum_{q=1}^\infty\left(\int_{x}^{y}\overline{\phi}_a(v,t)\phi_q(v,t)\, dv\right)\left(\int_{y}^{L_+}\phi_b(w,t)\overline{\phi}_q(w,t)\, dw\right)\nonumber\\
&\qquad\qquad\qquad\qquad +\xi\,\overline{\xi}\sum_{q=1}^\infty\left(\int_{y}^{L_+}\overline{\phi}_a(v,t)\phi_q(v,t)\, dv\right)\left(\int_{y}^{L_+}\phi_b(w,t)\overline{\phi}_q(w,t)\, dw\right)\, .
\end{align}
We will show first that the second term on the right hand sided of (\ref{i6}) cancels the fourth using the completeness of the $\phi_k(v)$ functions
\be
\int_{L_-}^{L_+}\overline{\phi}_a(v,t)\phi_b(v,t)=\delta_{a,b}\, ,\ \ \ \sum_{a=1}^\infty\overline{\phi}_a(x,t)\phi_a(y,t)=\delta(x-y)\, .
\ee
We introduce $\widetilde{\overline{\phi}}_a(v,t)=\mathbf{1}_{[y,L_+]}\overline{\phi}_a(v,t)$ and $\widetilde{\phi}_b(v,t)=\mathbf{1}_{[y,L_+]}\phi_b(v,t)$
with $\mathbf{1}_{[y,L_+]}$ the characteristic function of the interval $[y,L_+]$ i.e., is 1 when $v$ is inside the interval and 0 otherwise. We have
\begin{align}
\int_{y}^{L_+}\overline{\phi}_a(v,t)\phi_b(v,t)\, dv&=\int_{L_-}^{L_+} \widetilde{\overline{\phi}}_a(v,t)\widetilde{\phi}_b(v,t)\, dv\, ,\nonumber\\
&=\int_{L_-}^{L_+} \int_{L_-}^{L_+} \widetilde{\overline{\phi}}_a(v,t)\delta(v-w)\widetilde{\phi}_b(w,t)\, dv dw\, ,\nonumber\\
&=\sum_{q=1}^\infty\left(\int_{L_-}^{L_+}\widetilde{\overline{\phi}}_a(v,t)\phi_q(v,t)\, dv \right)\left(\int_{L_-}^{L_+}\overline{\phi}_q(w,t)\widetilde{\phi}_b(w,t)\, dw\right)\, , \nonumber\\
&=\sum_{q=1}^\infty\left(\int_{y}^{L_+}\overline{\phi}_a(v,t)\phi_q(v,t)\, dv \right)\left(\int_{y}^{L_+}\overline{\phi}_q(w,t)\phi_b(w,t)\, dw\right)\, .
\end{align}
This relation together with $\xi+\overline\xi=\xi\overline\xi=2-2\cos \pi\kappa$ shows that the second and the fourth term of (\ref{i6}) cancel. In a similar fashion
it can be shown that the third term is zero. Considering $\widetilde{\overline{\phi}}_a(v,t)=\mathbf{1}_{[x,y]}\overline{\phi}_a(v,t)$ and $\widetilde{\phi}_b(v,t)=\mathbf{1}_{[y,L_+]}\phi_a(v,t)$
we have $\int_{L_-}^{L_+} \widetilde{\overline{\phi}}_a(v,t)\widetilde{\phi}_b(v,t)\, dv=0$ and by inserting a resolution of the identity this is identical with the
third term on the r.h.s. of (\ref{i6}). Therefore, for $x<y$ we have
\be\label{i11}
\textsf{U}_{a,b}^{(-)}(x,t;y,t)-\delta_{a,b}=-\overline{\xi} \int_{x}^{y}\overline{\phi}_a(v,t)\phi_b(v,t)\, dv\, .
\ee
In the case $x>y$ we obtain
\be\label{i12}
\textsf{U}_{a,b}^{(-)}(x,t;y,t)-\delta_{a,b}=-\xi \int_{y}^{x}\overline{\phi}_a(v,t)\phi_b(v,t)\, dv\, .
\ee
From (\ref{i11}) and (\ref{i12}) we obtain (\ref{vsmall}). The result for the lattice case is given by (\ref{vsmalll}).

\twocolumngrid

\end{document}